\RequirePackage{fix-cm}
\documentclass[twocolumn]{svjour3}          
\smartqed  
\usepackage{natbib}
	\setcitestyle{aysep={}}
\usepackage{graphicx}
\usepackage{tabularx}
\usepackage{amssymb}
\usepackage{amsmath}
\usepackage[breaklinks]{hyperref}					
	\hypersetup{
    		colorlinks=true,
    		linkcolor=blue,
    		citecolor=blue,
    		filecolor=magenta,
    		urlcolor=blue}

\usepackage{nicefrac}					
\usepackage{makecell} 					
\usepackage{rotating}					
\usepackage{accents1}		
\newcommand*{\dt}[1]{%
  \accentset{\mbox{\large\bfseries .}}{#1}}
  
\begin{document}

\title{Time-Response Functions of Fractional-Derivative \\ Rheological Models}

\author{Nicos Makris         \and
        Eleftheria Efthymiou 
}


\institute{N. Makris \at
              Dept. of Civil and Environmental Engineering, \\
              Southern Methodist University, Dallas, Texas, 75276 \\
              Tel.: +1-(214)-768-1929\\
              \email{nmakris@smu.edu}   \\ \\
              Office of Theoretical and Applied Mechanics, \\
              Academy of Athens, Athens 10679, Greece \\
           \and
           E. Efthymiou \at
              Dept. of Civil and Environmental Engineering, Southern Methodist University, Dallas, Texas, 75276}

\date{Received: date / Accepted: date}

\maketitle

\begin{abstract}
In view of the increasing attention to the time responses of complex fluids described by power-laws in association with the need to capture inertia effects that manifest in high-frequency microrheology, we compute the five basic time-response functions of in-series or in-parallel connections of two elementary fractional derivative elements known as the Scott-Blair (springpot) element. The order of fractional differentiation in each Scott-Blair element is allowed to exceed unity reaching values up to 2 and at this limit-case the Scott-Blair element becomes an \textit{inerter} -- a mechanical analogue of the electric capacitor that its output force is proportional only to the relative acceleration of its end-nodes. With this generalization, inertia effects may be captured beyond the traditional viscoelastic behavior. In addition to the relaxation moduli and the creep compliances, we compute closed form expressions of the memory functions, impulse fluidities (impulse response functions) and impulse strain-rate response functions of the generalized fractional derivative Maxwell fluid, the generalized fractional derivative Kelvin-Voigt element and their special cases that have been implemented in the literature. Central to these calculations is the fractional derivative of the Dirac delta function which makes possible the extraction of singularities embedded in the fractional derivatives of the two-parameter Mittag-Leffler function that emerges invariably in the time-response functions of fractional derivative rheological models.
\end{abstract}

\section{Introduction}
Phenomenological constitutive models containing differential operators of non-integer order (fractional - \linebreak derivative models) have been proposed in mechanics, geosciences, electrical networks and biology over the last decades \citep[and references reported therein]{Gemant1936, Gemant1938, ScottBlair1944, ScottBlair1947, ScottBlairVeinoglouCaffyn1947, ScottBlairCaffyn1949, CaputoMainardi1971, Rabotnov1980,  BagleyTorvick1983a, BagleyTorvick1983b, Koeller1984, KohKelly1990,  Friedrich1991, GlockleNonnenmacher1991, GlockleNonnenmacher1994, MakrisConstantinou1991, SchiesselMetzlerBlumenNonnenmacher1995, Makris1997,  GorenfloMainardi1997, ChallamelZoricaAtanackovicSpasic2013, AtanackovicJanevOparnicaPilipovicZorica2015, WesterlundEkstam1994, LorenzoHartley2002, SukiBarabasiLutchen1994, PuigdeMoralesMarinkovicTurnerButlerFredbergSuresh2007}. Given that fractional derivative operators are linear differential operators, the time-dependent behavior of mechanical, electrical or biological networks can be computed with frequency-domain techniques in association with the Fourier or Laplace transforms \citep{LePage1961, Papoulis1962, Bracewell1965, Mainardi2010}.

There are several cases however, where the linear network that is described with a fractional derivative constitutive law is embedded in a wider system that behaves nonlinearly. As an example, seismic protection devices or rail pads which have been described with fractional derivative constitutive models \citep{KohKelly1990, MakrisConstantinou1991, Makris1992, ZhuCaiSpanos2015} belong to a structure or a vehicle-track system that may exhibit an overall nonlinear response. In this case the overall system response needs to be computed in the time-domain; therefore, a time-domain representation of the behavior of the individual components (devices) is needed. A time-domain representation is possible either by expressing the fractional derivatives with a time-series expansion, or by computing the basic time-response functions of the embedded linear networks and proceeding with an integral formulation to compute the overall system response.

When constitutive models with fractional-order \linebreak derivatives are involved the numerical evaluation of the fractional derivative of a  function is computationally demanding, partly because the fractional derivatives of the "through" and "across" variables of the linear network (stress, force, current = through variables and strain, displacement, voltage = across variables) need to be expressed via the Gr{\"u}nwald-Letnikov definition of the fractional derivative of order $q\in\mathbb{R}^+$ of a continuous function $f(t)$ \citep{OldhamSpanier1974, SamkoKilbasMarichev1974, MillerRoss1993, Podlubny1998}.
 \begin{align}\label{eq:Eq01}
&\frac{\mathrm{d}^qf(t)}{\mathrm{d}t^q}=D^qf(t_j) = \\ \nonumber
& \lim_{n\to\infty} \frac{\text{1}}{\Gamma(-q)}\left( \frac{t_j}{n} \right)^{-q} \sum\limits_{k=\text{0}}^{n-\text{1}} \frac{\Gamma(k-q)}{\Gamma(k+1)}f\left( t_j - k\frac{t_j}{n} \right)  \text{,} \enskip q\in\mathbb{R}^+
\end{align}
where $\mathbb{R}^+$ is the set of positive real numbers.

The Gr{\"u}nwald-Letnikov definition given by Eq. \eqref{eq:Eq01} indicates that a large number of terms ($n \rightarrow \infty$) may be needed to meet satisfactory convergence and the computational effort may be intense \citep{Makris1992, MillerRoss1993}. Accordingly, an integral formulation after deriving the basic time-response functions of the embedded linear networks that involve fractional differential operators emerge as an attractive alternative. Expressions of the relaxation modulus and the creep compliance of selective fractional derivative viscoelastic models have been presented by \citet{SmitDeVries1970, Koeller1984, Friedrich1991, GlockleNonnenmacher1991, GlockleNonnenmacher1994, HeymansBauwens1994,  SukiBarabasiLutchen1994, SchiesselMetzlerBlumenNonnenmacher1995, PaladeVerneyAttane1996, DjordjevicJaricFrabryFredbergStamenovic2003, CraiemMagin2010, Mainardi2010, MainardiSpada2011, Hristov2019}. The present work builds upon the aforementioned studies and constructs additional time-re\-sponse functions such as the memory function, the impulse fluidity (impulse response function) and the impulse strain-rate response function of the generalized fractional Maxwell fluid and the generalized fractional Kelvin-Voigt element which are in series or in parallel connections of two elementary fractional-derivative elements known as the Scott-Blair (springpot) model \citep{ScottBlair1944, ScottBlair1947}. Special cases of these generalized fractional-derivative models are the spring -- Scott-Blair in-series or parallel connections that have been used by \citet{SukiBarabasiLutchen1994} to express the pressure--volume relation of the lung tissue viscoelastic behavior and subsequently used by \citet{PuigdeMoralesMarinkovicTurnerButlerFredbergSuresh2007} to model the viscoleastic behavior of human red blood cells. The springpot -- dashpot in-series or parallel connections are rheological models have been used to capture the high-frequency behavior of semiflexible polymer networks \citep{GittesMacKintosh1998, AtakhorramiMizunoKoenderinkLiverpoolMacKintoshSchmidt2008, DominguezGarciaCardinauxBertsevaForroScheffoldJeney2014} or the behavior of viscoelastic dampers for the vibration and seismic isolation of structures \citep{Makris1992, MakrisConstantinou1992, MakrisDeoskar1996}.

The memory function of the elementary Scott-Blair (springpot) element is central in this work, since it is the fractional derivative of the Dirac delta function which is merely the kernel appearing in the convolution of the Riemann-Liouville definition of the fractional derivative of a function. This finding shows that the fractional derivative of the Dirac delta function is finite everywhere other than at the singularity point and it is the inverse Fourier transform of $(\operatorname{i}\omega)^q$ with $q\in \mathbb{R}^+$. It emerges as a key function in the derivation of the time-response functions of generalized fractional derivative rheological models, since it makes possible the extraction of the singularities embedded in the fractional derivatives of the two-parameter Mittag-Leffler function.

\section{Basic time-response functions of linear phenomenological models}
This paper studies the integral representation of linear phenomenological constitutive models (linear networks) of the form
\begin{equation}\label{eq:Eq02}
\left[ \sum\limits_{m=\text{0}}^{M} a_m  \frac{\mathrm{d}^{p_m}}{\mathrm{d}t^{p_m}} \right] \tau(t) =\left[ \sum\limits_{n=\text{0}}^{N} b_n  \frac{\mathrm{d}^{q_n}}{\mathrm{d}t^{q_n}} \right] \gamma(t)
\end{equation}
where $\tau(t)$ and $\gamma(t)$ are the time-histories of the stress and the small-gradient strain, $a_m$ and $b_n$ are real-valued frequency-independent coefficients and the order of differentiation, $p_m$ and $q_n$ are real, positive non-integer numbers (usually rational fractions). A definition of the fractional derivative of order $q$ is given through the convolution integral
\begin{equation}\label{eq:Eq03}
I^q\gamma(t)=\frac{\text{1}}{\Gamma(q)}\int_{c}^{t}\gamma(\xi)(t-\xi)^{q-\text{1}}\mathrm{d}\xi
\end{equation}
where $\Gamma(q)$ is the Gamma function. When the lower limit, $c=\text{0}$, the integral given by Eq. \eqref{eq:Eq03} is often referred to as the Riemann-Liouville fractional integral \citep{OldhamSpanier1974, SamkoKilbasMarichev1974, MillerRoss1993, Podlubny1998}. The above integral converges only for $q>\text{0}$, or in the case where $q$ is a complex number, the integral converges for $\mathcal{R}(q)>\text{0}$. Nevertheless, by a proper analytic continuation across the line $\mathcal{R}(q)=\text{0}$, and provided that the function $f(t)$ in $n$ times differentiable, it can be shown that the integral given by Eq. \eqref{eq:Eq03} exists for $n-\mathbb{R}(q)>\text{0}$ \citep{Riesz1949}. In this case the generalized (fractional) derivative of order $q\in \mathbb{R}^+$ exists and is defined as
\begin{equation}\label{eq:Eq04}
\frac{\mathrm{d}^q\gamma(t)}{\mathrm{d}t^q}=I^{-q}\gamma(t)=\frac{\text{1}}{\Gamma(-q)}\int_{\text{0}^-}^{t}\frac{\gamma(\xi)}{(t-\xi)^{q+\text{1}}} \mathrm{d}\xi  \text{,} \enskip q\in \mathbb{R}^+
\end{equation}
where $\mathbb{R}^+$ is the set of positive real numbers and the lower limit of integration, $\text{0}^-$, may engage an entire singular function at the time origin such as $\gamma(t)=\delta(t-\text{0})$ \citep{Lighthill1958, Mainardi2010}. Eq. \eqref{eq:Eq04} indicates that the fractional derivative of order $q$ of $\gamma(t)$ is essentially the convolution of $\gamma(t)$ with the kernel {\large $\nicefrac{t^{-q-\text{1}}}{\Gamma(-q)}$} \citep{OldhamSpanier1974, SamkoKilbasMarichev1974, MillerRoss1993, Mainardi2010}. The Rie\-mann-Liouville definition of the fractional derivative of order $q\in \mathbb{R}^+$ given by Eq. \eqref{eq:Eq04}, where the lower limit of integration is zero, is central in this work since the strain and stress histories, $\gamma(t)$ and $\tau(t)$, are causal functions, being zero at negative times.

Linear viscoelastic materials, such as those described with Eq. \eqref{eq:Eq02}, obey the so-called Boltzmann superposition principle -- that the output response history can be obtained as the convolution of the input history after being convoluted with the corresponding time-response function. The basic time-response functions can be obtained either by imposing an impulse or a unit-step excitation on the constitutive model, or by inverting in the time-domain the corresponding frequency-response functions of the real-parameter constitutive model. Such techniques are well known in the literature of rheology \citep{Ferry1980, BirdArmstrongHassager1987, Tschoegl1989}, structural mechanics \citep{HarrisCrede1976, VeletsosVerbic1974, Makris1997b} and automatic control \citep{Bode1945, Reid1983, TriverioGrivetNakhlaCanaveroAchar2007}.

The linearity of Eq. \eqref{eq:Eq02} permits its transformation in the frequency domain
\begin{equation}\label{eq:Eq05}
\tau(\omega)\left[ \sum\limits_{m=\text{0}}^{M} a_m  (\operatorname{i}\omega)^{p_m} \right] =\gamma(\omega)\left[ \sum\limits_{n=\text{0}}^{N} b_n  (\operatorname{i}\omega)^{q_n} \right]
\end{equation}
where, {\small $\operatorname{i}=\sqrt{-\text{1}}=$} imaginary unit, {\small $\tau(\omega)=\int_{-\infty}^{\infty}\tau(t) e^{-\operatorname{i}\omega t}\mathrm{d}t$} and $\gamma(\omega)=\int_{-\infty}^{\infty}\gamma(t) e^{-\operatorname{i}\omega t}\mathrm{d}t$ are the Fourier transforms of the stress and strain histories and $(\operatorname{i}\omega)^{q}\gamma(\omega)$ is the Fourier Transform of the fractional derivative of order $q$ of the time function, $\gamma(t)$ \citep{OldhamSpanier1974, KohKelly1990, Mainardi2010, SamkoKilbasMarichev1974, MillerRoss1993, Podlubny1998}
\begin{equation}\label{eq:Eq06}
\mathcal{F}\left\lbrace \frac{\mathrm{d}^q\gamma(t)}{\mathrm{d}t^q} \right\rbrace = \int_{-\infty}^{\infty}\frac{\mathrm{d}^q\gamma(t)}{\mathrm{d}t^q} e^{-\operatorname{i}\omega t} \mathrm{d}t = (\operatorname{i}\omega)^{q}\gamma(\omega)
\end{equation}
Eq. \eqref{eq:Eq05} is expressed as 
\begin{equation}\label{eq:Eq07}
\tau(\omega)=\left[ G_{\text{1}}(\omega) + \operatorname{i}G_{\text{2}}(\omega) \right]\gamma(\omega)
\end{equation}
where $\mathcal{G}(\omega)= G_{\text{1}}(\omega) + \operatorname{i}G_{\text{2}}(\omega)]$ is the complex dynamic modulus of the constitutive model \citep{Ferry1980, BirdArmstrongHassager1987, Giesekus1995}.
\begin{equation}\label{eq:Eq08}
\mathcal{G}(\omega)=G_{\text{1}}(\omega) + \operatorname{i}G_{\text{2}}(\omega)= \frac{\sum\limits_{n=\text{0}}^{N} b_n  (\operatorname{i}\omega)^{q_n}}{ \sum\limits_{m=\text{0}}^{M} a_m  (\operatorname{i}\omega)^{p_m}}
\end{equation}
and is a frequency-response function that relates a stress output to a strain input. The stress, $\tau(t)$, in Eq. \eqref{eq:Eq02} can be computed in the time domain with the convolution integral
\begin{equation}\label{eq:Eq09}
\tau(t) = \int_{\text{0}}^tM(t-\xi)\gamma(\xi) \mathrm{d}\xi
\end{equation}
where $M(t-\xi)$ is the memory function of the model \citep{BirdArmstrongHassager1987, VeletsosVerbic1974, Makris1997b}, defined as the resulting stress at time $t$ due to an impulsive strain input at time $\xi$ ($\xi<t$), and is the inverse Fourier transform of the complex dynamic modulus 
\begin{equation}\label{eq:Eq10}
M(t)=\frac{\text{1}}{\text{2}\pi}\int_{-\infty}^{\infty}\mathcal{G}(\omega)e^{\operatorname{i}\omega t}\mathrm{d}\omega
\end{equation}
The inverse of the complex dynamic modulus is the complex dynamic compliance \cite{Pipkin1986, Giesekus1995}
\begin{equation}\label{eq:Eq11}
\mathcal{J}(\omega)=J_{\text{1}}(\omega)+\operatorname{i} J_{\text{2}}(\omega) = \frac{\text{1}}{\mathcal{G}(\omega)}=\frac{ \sum\limits_{m=\text{0}}^{M} a_m  (\operatorname{i}\omega)^{p_m}}{\sum\limits_{n=\text{0}}^{N} b_n  (\operatorname{i}\omega)^{q_n}}
\end{equation}
which is a frequency-response function that relates a strain output to a stress input. In structural mechanics the equivalent of the complex dynamic compliance is known as the dynamic flexibility, often expressed with $\mathcal{H}(\omega)$ \citep{CloughPenzien1970}. Accordingly, the strain history in Eq. \eqref{eq:Eq02} can be computed in the time domain via a convolution integral
\begin{equation}\label{eq:Eq12}
\gamma(t)=\int_{\text{0}}^t \phi(t-\xi)\tau(\xi) \mathrm{d}\xi
\end{equation}
where $\phi(t-\xi)$ is the impulse fluidity \citep{Giesekus1995}, defined as the resulting strain history at time $t$ due to an impulsive stress input at time $\xi(t<\xi)$, and is the inverse Fourier transform of the dynamic compliance, $\mathcal{J}(\omega)$ 
\begin{equation}\label{eq:Eq13}
\phi(t)=\frac{\text{1}}{\text{2}\pi}\int_{-\infty}^{\infty}\mathcal{J}(\omega)e^{\operatorname{i}\omega t} \mathrm{d}\omega
\end{equation}
In structural mechanics, the equivalent of the impulse fluidity at the force--displacement level is known as the impulse response function, $h(t)$, which is the kernel appearing in the Duhamel integral \citep{CloughPenzien1970, VeletsosVerbic1974, Makris1997b}. Expressions of the impulse fluidity of the Hookean solid, the Newtonian fluid, the Kelvin-Voigt solid and the Maxwell fluid have been presented by \citet{Giesekus1995}; whereas, for the three-parameter Poyinting-Thomson solid and the three- \linebreak[0] parameter Jeffreys fluid have been presented by \citet{MakrisKampas2009}.

\begin{table*}
\begin{center}
\caption{Basic frequency-response functions and their corresponding causal time-response functions in linear viscoelasticity and linear network analysis. Because the time-response functions listed in the right column are zero at negative times, their Fourier transform shown in the left column is also a Laplace transform with variable $s=\operatorname{i}\omega$}
\setlength\tabcolsep{0pt} 
\begin{tabular}{c}
{\centering \includegraphics[width=.95\textwidth, angle=0]{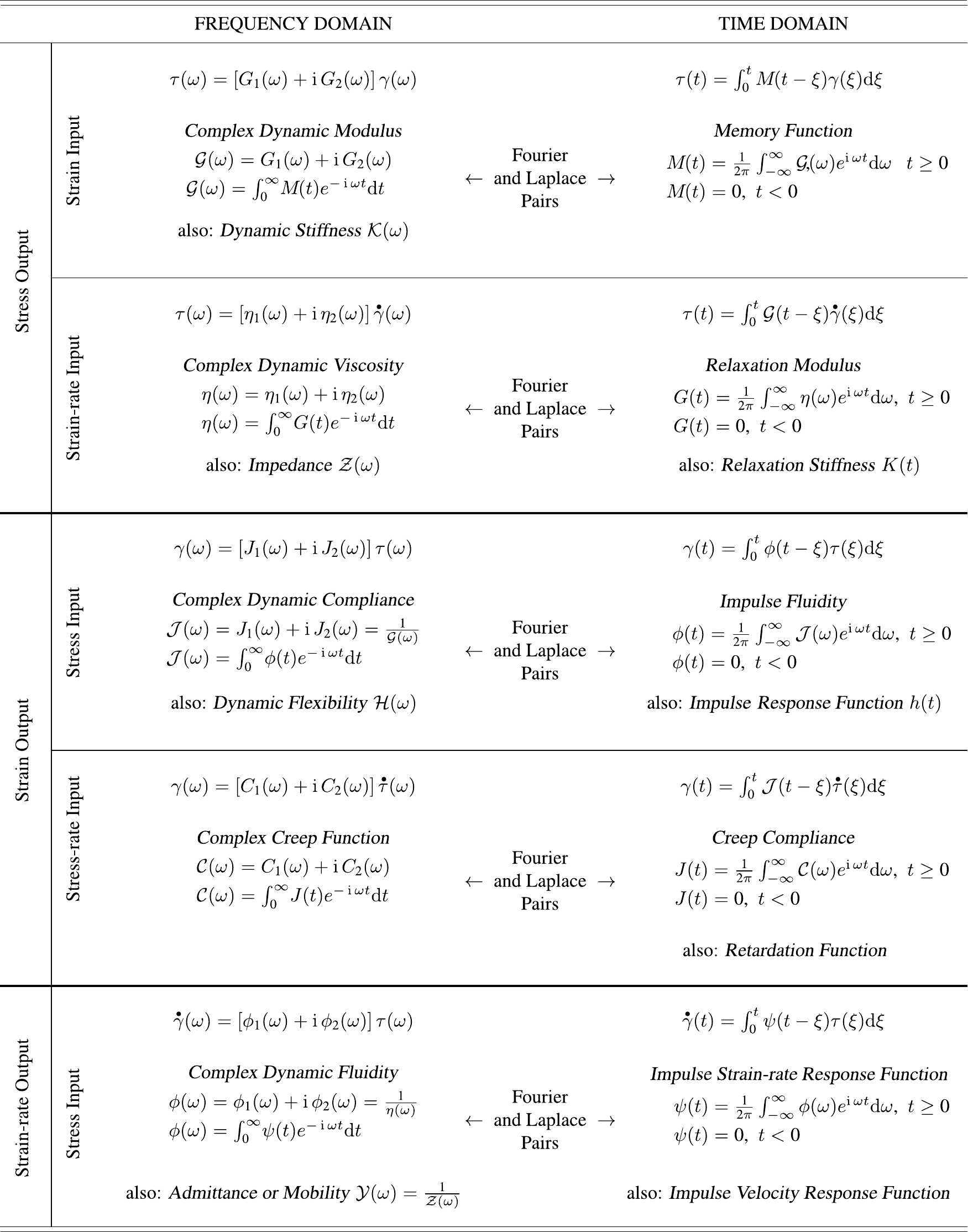}}
\end{tabular}
\label{tab:Table1}
\end{center}
\end{table*}

Another useful frequency-response function of a linear constitutive model is the complex dynamic viscosity, $\eta(\omega)=\eta_{\text{1}}(\omega)+\operatorname{i}\eta_{\text{2}}(\omega)$, which relates a stress output to a strain-rate input
\begin{equation}\label{eq:Eq14}
\tau(\omega)=\left[ \eta_{\text{1}}(\omega)+\operatorname{i}\eta_{\text{2}}(\omega) \right]\dt{\gamma}(\omega)
\end{equation} 
 where $\dt{\gamma}(\omega)=\operatorname{i}\omega \gamma(\omega)$ is the Fourier transform of the strain-rate history. In structural mechanics, the equivalent of the complex dynamic viscosity at the force-velocity level is known as the mechanical impedance, $\mathcal{Z}(\omega)=Z_{\text{1}}(\omega)+\operatorname{i}Z_{\text{2}}(\omega)$ \citep{HarrisCrede1976}. The term impedance and its notation, $\mathcal{Z}(\omega)$, have also been used to express the pressure--volume-rate relation of the lung tissue viscoelastic behavior of human and selective animal lungs \citep{SukiBarabasiLutchen1994}. For the linear viscoelastic model given by Eq. \eqref{eq:Eq02}, the complex dynamic viscosity (impedance) of the model is 
\begin{equation}\label{eq:Eq15}
\eta(\omega)=\eta_{\text{1}}(\omega)+\operatorname{i}\eta_{\text{2}}(\omega)=\frac{\sum\limits_{n=\text{0}}^{N} b_n  (\operatorname{i}\omega)^{q_n}}{ \sum\limits_{m=\text{0}}^{M} a_m  (\operatorname{i}\omega)^{p_m+\text{1}}}
\end{equation}

The stress $\tau(t)$ in Eq. \eqref{eq:Eq02} can be computed in the time domain with an alternative convolution integral 
\begin{equation}\label{eq:Eq16}
\tau(t)=\int_{\text{0}}^t G(t-\xi)\frac{\mathrm{d}\gamma(\xi)}{\mathrm{d}\xi}\mathrm{d}\xi
\end{equation}
where $G(t-\xi)$ is the relaxation modulus of the constitutive model defined as the resulting stress at the present time, $t$, due to a unit-step strain at time $\xi\text{ } (\xi<t)$ and is the inverse Fourier transform of the complex dynamic viscosity
\begin{equation}\label{eq:Eq17}
G(t)=\frac{\text{1}}{\text{2}\pi}\int_{-\infty}^{\infty} \eta(\omega) e^{\operatorname{i} \omega t} \mathrm{d}\omega
\end{equation}

Expressions for the relaxation modulus, $G(t)$, of the various simple viscoelastic models are well-known in the literature \citep{Ferry1980, BirdArmstrongHassager1987, Tschoegl1989, Giesekus1995}. Expressions for the relaxation modulus of simple fractional derivative viscoelastic models have been presented by \citet{SmitDeVries1970, Koeller1984, Friedrich1991, GlockleNonnenmacher1991, GlockleNonnenmacher1994, SukiBarabasiLutchen1994, SchiesselMetzlerBlumenNonnenmacher1995, PaladeVerneyAttane1996,  DjordjevicJaricFrabryFredbergStamenovic2003, PuigdeMoralesMarinkovicTurnerButlerFredbergSuresh2007, Mainardi2010, CraiemMagin2010, MainardiSpada2011, JaishankarMcKinley2013, Hristov2019}.

The inverse of the complex dynamic viscosity is the complex dynamic fluidity \citep{Giesekus1995}
\begin{equation}\label{eq:Eq18}
\phi(\omega)=\phi_{\text{1}}(\omega)+\operatorname{i}\phi_{\text{2}}(\omega)=\frac{\text{1}}{\eta(\omega)}=\frac{ \sum\limits_{m=\text{0}}^{M} a_m  (\operatorname{i}\omega)^{p_m+\text{1}}}{\sum\limits_{n=\text{0}}^{N} b_n  (\operatorname{i}\omega)^{q_n}}
\end{equation}
which is a frequency-response  function that relates a strain-rate output to a stress input. In structural mechanics the equivalent of the complex dynamic fluidity at the velocity-force level is known as the mechanical admittance or mobility \citep{HarrisCrede1976}. The strain-rate history, $\dt{\gamma}(t)$, can be computed in the time domain via the convolution integral
\begin{equation}\label{eq:Eq19}
\dt{\gamma}(t)=\frac{\mathrm{d}\gamma(t)}{\mathrm{d}t} =\int_{\text{0}}^{t} \psi(t-\xi) \tau(\xi) \mathrm{d}\xi
\end{equation}
where $\psi(t-\xi)$ is the impulse strain-rate response function defined as the resulting strain-rate output at time $t$ due to an impulsive stress input at time $\xi(\xi<t)$ and is the inverse Fourier  transform of the dynamic fluidity
\begin{equation}\label{eq:Eq20}
\psi(t)=\frac{\text{1}}{\text{2}\pi}\int_{-\infty}^{\infty} \phi(\omega) e^{\operatorname{i} \omega t} \mathrm{d}\omega
\end{equation}

Together with the relaxation modulus, $G(t-\xi)$ that appears as a kernel in Eq. \eqref{eq:Eq16}, the other most popular time-response function in experimental stress analysis is the creep compliance, $J(t-\xi)$ \citep{Ferry1980, Pipkin1986, Tschoegl1989, BirdArmstrongHassager1987}, that is defined as the resulting strain, $\gamma(t)$, at the present time $t$ due to a unit-step stress at time $\xi(\xi<t)$ and is the inverse Fourier transform of the complex creep function, $\mathcal{C}(\omega)$
\begin{equation}\label{eq:Eq21}
J(t)=\frac{\text{1}}{\text{2}\pi}\int_{-\infty}^{\infty} \mathcal{C}(\omega) e^{\operatorname{i} \omega t} \mathrm{d}\omega
\end{equation}
The complex creep function, $\mathcal{C}(\omega)$, is the ratio of the cyclic strain output $\gamma(\omega)$, over the cyclic stress-rate input $\dt{\tau}(\omega)$ \citep{MasonGanesanVanZantenWirtzKuo1997, Mason2000, EvansTassieriAuhlWaigh2009, Makris2019}. 
\begin{equation}\label{eq:Eq22}
\mathcal{C}(\omega) = \frac{\gamma(\omega)}{\dt{\tau}(\omega)}=\frac{\text{1}}{\operatorname{i}\omega}\mathcal{J}(\omega)
\end{equation}
Under a unit-amplitude step-stress, $\tau(t)=U(t-\text{0})$ where $U(t-\text{0})$ is the Heaviside unit-step function, the strain history, $\gamma(t)$ is
\begin{equation}\label{eq:Eq23}
\gamma(t)=\int_{\text{0}}^t J(t-\xi)\frac{\mathrm{d}\tau(\xi)}{\mathrm{d}\xi}\mathrm{d}\xi
\end{equation}
All five time-response functions given by Eqs. \eqref{eq:Eq10}, \eqref{eq:Eq13}, \eqref{eq:Eq17} and \eqref{eq:Eq21} are causal time-response functions -- that is they are zero at negative times. This means that their Fourier transform vanishes at negative times and it becomes one-sided. For instance, the complex dynamic viscosity, $\eta(\omega)$, that is the Fourier transform of the relaxation modulus, $G(t)$, is
\begin{equation}\label{eq:Eq24}
\eta(\omega)=\int_{-\infty}^{\infty} G(t)e^{-\operatorname{i}\omega t}\mathrm{d}t=\int_{\text{0}}^{\infty} G(t)e^{-\operatorname{i}\omega t}\mathrm{d}t
\end{equation}
The one-sided integral on the right-hand side of Eq. \eqref{eq:Eq24} that results from the causality of the time-response function, ($G(t)=\text{0}$ when $t<\text{0}$), is essentially the Laplace transform of the time-response function \citep{LePage1961}
\begin{equation}\label{eq:Eq25}
\eta (s)=\mathcal{L}\left\lbrace G(t) \right\rbrace =\int_{\text{0}}^{\infty}G(t)e^{-s t} \mathrm{d}t
\end{equation}
where $s=\operatorname{i}\omega$ is the Laplace variable and $\mathcal{L}$ indicates the Laplace transform operator. Accordingly, the frequency-response functions given by Eqs. \eqref{eq:Eq08}, \eqref{eq:Eq11}, \eqref{eq:Eq15}, \eqref{eq:Eq18} and \eqref{eq:Eq22} are Laplace pairs with their corresponding time-response functions given by Eqs. \eqref{eq:Eq10}, \eqref{eq:Eq13}, \eqref{eq:Eq17} and \eqref{eq:Eq21}, which are summarized in Table \ref{tab:Table1} when a strain input, strain-rate input, stress input or stress-rate input is imposed.

\section{The fractional derivative of the Dirac delta function}
Following the observations by \citet{Nutting1921}, that the stress response of several fluid-like materials to a step-strain decays following a power law $(\tau(t)=G(t)\sim t^{-q}$ with $\text{0}<q<\text{1})$ and the early work of \citet{Gemant1936, Gemant1938} on fractional differentials; \citet{ScottBlair1944, ScottBlair1947} pioneered the introduction of fractional calculus in viscoelasticity. With analogy to the Hookean spring, in which the stress is proportional to the zero-th derivative of the strain and the Newtonian dashpot, in which the stress is proportional to the first derivative of the strain, \citet{ScottBlair1944, ScottBlair1947, ScottBlairVeinoglouCaffyn1947, ScottBlairCaffyn1949} proposed the \textit{springpot} element -- that is an element in-between a spring and a dashpot with constitutive law
\begin{equation}\label{eq:Eq26}
\tau(t)=K_q\frac{\mathrm{d}^q\gamma(t)}{\mathrm{d}t^q}
\end{equation} 
where $q$ is a positive real number, 0 $\leq q \leq$ 1, $K_q$ is a phenomenological material parameter with units [M][L]$^{-\text{1}}$[T]$^{q-\text{2}}$ (say Pa-sec$^q$) and {\large$\nicefrac{\mathrm{d}^q\gamma(t)}{\mathrm{d}t^q}$} is the fractional derivative of the strain-history defined by Eq. \eqref{eq:Eq04}.

For the elastic Hookean spring with elastic modulus, $G$, its memory function as defined by Eq. \eqref{eq:Eq10} is $M(t)=G\delta(t-\text{0})$ -- that is the zero-order derivative of the Dirac delta function; whereas, for the Newtonian dashpot with viscosity, $\eta$,  its memory function is $M(t)=\eta \,${\large$\frac{\mathrm{d}\delta(t-\text{0})}{\mathrm{d}t}$} -- that is the first-order derivative of the Dirac delta function \citep[see also Table \ref{tab:Table2}]{BirdArmstrongHassager1987}. Since the springpot element defined by Eq. \eqref{eq:Eq26} with 0 $\leq q \leq$ 1 is a constitutive model that is in-between the Hookean spring and the Newtonian dashpot, physical continuity suggests that the memory function of the springpot model given by Eq. \eqref{eq:Eq26} shall be of the form of $M(t)=K_q \,$ {\large$\frac{\mathrm{d}^q\delta(t-\text{0})}{\mathrm{d}t^q}$} -- that is the fractional derivative of order $q$ of the Dirac delta function \citep{OldhamSpanier1974, Podlubny1998}.

\begin{figure}[t!]
\centering
\includegraphics[width=0.47\textwidth, angle=0]{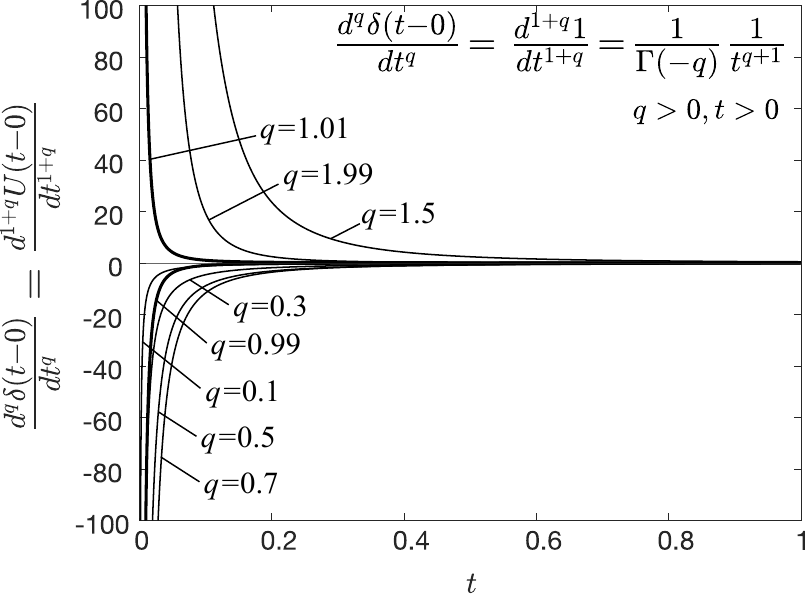}
\caption{Plots of the fractional derivative of the Dirac delta function of order $q\in\left\lbrace\mathbb{R}^+-\mathbb{N}\right\rbrace$, which are the $\text{1}+q$ order derivative of the constant 1 for positive times. The functions are finite everywhere other than the time origin, $t=\text{0}$. Figure \ref{fig:Fig01} shows that the fractional derivatives of the singular Dirac delta function and these of the constant unit at positive times are expressed with the same family of functions.}
\label{fig:Fig01}
\end{figure}

The fractional derivative of the Dirac delta function emerges directly from the property of the Dirac delta function \citep{Lighthill1958}
\begin{equation}\label{eq:Eq27}
\int_{-\infty}^{\infty}\delta(t-\xi)f(t)\mathrm{d}t=f(\xi)
\end{equation}
By following the Riemann-Liouville definition of the fractional derivative of a function given by the convolution appearing in Eq. \eqref{eq:Eq04}, the fractional derivative of order $q\in \mathbb{R}^+$ of the Dirac delta function is
\begin{equation}\label{eq:Eq28}
\frac{\mathrm{d}^q\delta(t-\xi)}{\mathrm{d}t^q}=\frac{\text{1}}{\Gamma(-q)}\int_{\text{0}^-}^t \frac{\delta(\tau-\xi)}{(t-\tau)^{\text{1}+q}}\mathrm{d}\tau \text{,} \enskip q\in \mathbb{R}^+
\end{equation} 
and by applying the property of the Dirac delta function given by Eq. \eqref{eq:Eq27}; Eq. \eqref{eq:Eq28} gives
\begin{equation}\label{eq:Eq29}
\frac{\mathrm{d}^q\delta(t-\xi)}{\mathrm{d}t^q}=\frac{\text{1}}{\Gamma(-q)}\frac{\text{1}}{(t-\xi)^{\text{1}+q}}  \text{,} \enskip q\in \mathbb{R}^+
\end{equation}
Eq. \eqref{eq:Eq29} offers the remarkable result that the fractional derivative of the Dirac delta function of any order $q\in\left\lbrace\mathbb{R}^+-\mathbb{N}\right\rbrace$ is finite everywhere other than at $t=\xi$; whereas, the Dirac delta function and its integer-order derivatives are infinite-valued, singular functions that are understood as a monopole, dipole and so on; and we can only interpret them through their mathematical properties as the one given by Eq. \eqref{eq:Eq27}. Figure \ref{fig:Fig01} plots the fractional derivative of the Dirac delta function at $\xi=\text{0}$
\begin{equation}\label{eq:Eq30}
\frac{\mathrm{d}^q\delta(t-\text{0})}{\mathrm{d}t^q}=\frac{\text{1}}{\Gamma(-q)}\frac{\text{1}}{t^{\text{1}+q}} \enskip \enskip \text{with} \enskip  q\in \mathbb{R}^+ \text{,} \enskip t>\text{0}
\end{equation}
The result of Eq. \eqref{eq:Eq30} is identical to the alternative definition of the $n^{\text{th}}$ $(n \in \mathbb{N}_{\text{0}})$ derivative of the Dirac delta function presented by \citet{GelfandShilov1964} with a proper interpretation of the quotien {\large $\frac{\text{1}}{t^{n+\text{1}}}$} as a limit at $t=\text{0}$.
\begin{equation}\label{eq:Eq31}
\frac{\mathrm{d}^n\delta(t-\text{0})}{\mathrm{d}t^n}=\frac{\text{1}}{\Gamma(-n)}\frac{\text{1}}{t^{n+\text{1}}} \enskip \text{with} \enskip n \in \mathbb{N}_{\text{0}}
\end{equation}
where $\mathbb{N}_{\text{0}}$ is the set of positive integers including zero

The result for the fractional derivative of the Dirac delta function given by Eq. \eqref{eq:Eq30} is also compared with the well known results in the literature for the fractional derivative of the constant unit function $f(t)=\text{1}$ \citep{OldhamSpanier1974, SamkoKilbasMarichev1974,  MillerRoss1993, Podlubny1998}.
\begin{equation}\label{eq:Eq32}
D^r\text{1}=\frac{t^{-r}}{\Gamma (\text{1}-r)} \text{,} \enskip r\in \mathbb{R}^+ \enskip \text{and} \enskip t>\text{0}
\end{equation}
For $ r\in \mathbb{N}$, $D^r\text{1}=\text{0}$ due to the poles of the Gamma function at 0, -1, -2 and the classical results are recovered. Clearly, in Eq. \eqref{eq:Eq32} time needs to be positive ($t>\text{0}$); otherwise, the result of Eq. \eqref{eq:Eq32} would be a complex number when $r\in\left\lbrace\mathbb{R}^+-\mathbb{N}\right\rbrace$. Accordingly, a more formal expression of equation Eq. \eqref{eq:Eq32} within the context of generalized functions is
\begin{equation}\label{eq:Eq33}
D^r U(t-\text{0})=\frac{\text{1}}{\Gamma(\text{1}-r)}\frac{\text{1}}{t^r}\text{,} \enskip r\in\mathbb{R}^+ \text{,} \enskip t>\text{0}
\end{equation}
where $U(t-\text{0})$ is the Heaviside unit-step function at the time origin \citep{Lighthill1958}.

For the case where $r>\text{1}$, $\text{1}-r=-q$ with $q\in \mathbb{R}^+$; therefore $\text{1}+q=r>\text{1}$. Accordingly, for $r>\text{1}$, Eq. \eqref{eq:Eq33} can be expressed as
\begin{align}\label{eq:Eq34}
\frac{\mathrm{d}^{\text{1}+q}}{\mathrm{d}t^{\text{1}+q}}U(t-\text{0}) & = \frac{\mathrm{d}^q}{\mathrm{d}t^q}\left[ \frac{\mathrm{d}^{\text{1}}U(t-\text{0})}{\mathrm{d}t} \right] = \\ \nonumber
\frac{\mathrm{d}^q}{\mathrm{d}t^q} \delta(t-\text{0}) & =\frac{\text{1}}{\Gamma(-q)}\frac{\text{1}}{t^{\text{1}+q}} \text{,} \enskip q\in \mathbb{R}^+\text{,} \enskip t>\text{0}
\end{align}
and the result of Eq. \eqref{eq:Eq30} is recovered. In Eq. \eqref{eq:Eq34} we used that $\delta(t-\text{0})=$ {\large$\frac{\mathrm{d}^{\text{1}}U(t-\text{0})}{\mathrm{d}t}$} \citep{Lighthill1958}.

\section{Time-response functions of the Scott-Blair (springpot) element}
The memory function, $M(t)$ appearing in Eq. \eqref{eq:Eq09}, of the Scott-Blair $($springpot when 0 $\leq q \leq\text{1})$ element expressed by Eq. \eqref{eq:Eq26} results directly from the definition of the fractional derivative expressed with the Reimann-Liouville integral given by Eq. \eqref{eq:Eq04}. Substitution of Eq. \eqref{eq:Eq04} into Eq. \eqref{eq:Eq26} gives
\begin{equation}\label{eq:Eq35}
\tau(t)=\frac{K_q}{\Gamma(-q)}\int_{\text{0}}^t \frac{\gamma(\xi)}{(t-\xi)^{q+\text{1}}} \mathrm{d}\xi \text{,} \enskip q\in \mathbb{R}^+
\end{equation} 
By comparing Eq. \eqref{eq:Eq35} with Eq. \eqref{eq:Eq09}, the memory function, $M(t)$, of the Scott-Blair (springpot when 0 $\leq q \leq$ 1) element is merely the kernel of the Riemann-Liouville convolution multiplied with the material parameter $K_q$
\begin{equation}\label{eq:Eq36}
M(t)=\frac{K_q}{\Gamma(-q)}\frac{\text{1}}{t^{q+\text{1}}}=K_q\frac{\mathrm{d}^q\delta(t-\text{0})}{\mathrm{d}t^q}\text{,} \enskip q\in \mathbb{R}^+
\end{equation}
where the right-hand side of Eq. \eqref{eq:Eq36} is from Eq. \eqref{eq:Eq30}. Eq. \eqref{eq:Eq36} shows that the memory function of the springpot element is the fractional derivative of order $q\in \mathbb{R}^+$ of the Dirac delta function as was anticipated by using the argument of physical continuity given that the springpot element interpolates the Hookean spring and the Newtonian dashpot.

In this study we adopt the name ``Scott-Blair element'' rather than the more restrictive ``springpot'' element given that the fractional order of differentiation $q\in \mathbb{R}^+$ is allowed to take values larger than one. For instance when $\text{1}\leq q \leq \text{2}$ the Scott-Blair element represents an element that is in-between a dashpot and an inerter.

An ``inerter'' is a linear mechanical element where at the force-displacement level the output force is proportional only to the relative acceleration of its end-nodes (terminals) \citep{Smith2002, PapageorgiouSmith2005, MakrisMoghimi2018} and complements the classical elastic spring and viscous dashpot. In a stress--current/strain-rate--voltage analogy, the inerter is the mechanical analogue of the electric capacitor and its constant of proportionality is the distributed inertance, $m_R$,  with units of $\left[ \text{M} \right]\left[ \text{L} \right]^{-\text{1}}$ (say \textsl{Pa-sec}$^{\text{2}}$). Studies in high-frequency microrheology \citep{MasonWeitz1995, IndeiSchieberCordoba2012, DominguezGarciaCardinauxBertsevaForroScheffoldJeney2014} account for the distributed inertance $m_R=$ {\large $\frac{m}{\text{6}\pi R}$}, where $m$ and $R$ are the mass and radius of the \noindent particle (bead), respectively. When $q=\text{2}$, the memory function of the Scott-Blair element given by Eq. \eqref{eq:Eq36} gives
\begin{figure*}[t!]
\centering
\includegraphics[width=.72\textwidth, angle=0]{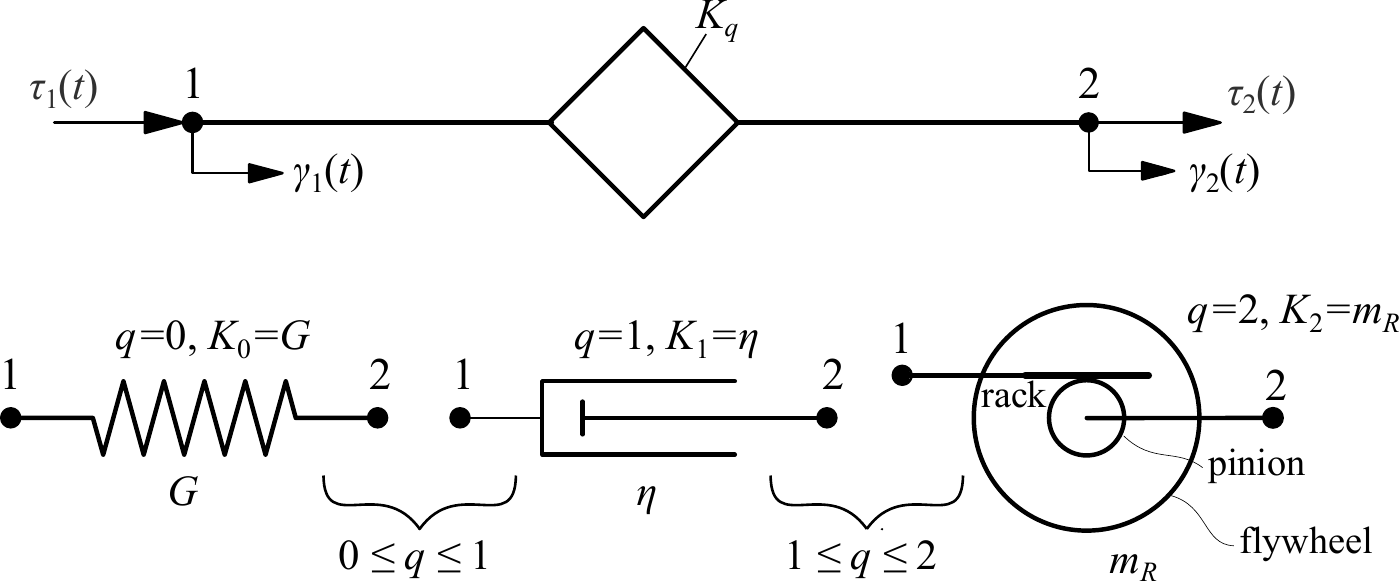}
\caption{The Scott-Blair element with constitutive law $\tau(t)=K_q${\large $\nicefrac{\mathrm{d}^q\gamma(t)}{\mathrm{d}t^q}$} is an element in-between a spring ($K_{\text{0}}=G$) and a dashpot ($K_{\text{1}}=\eta$) when $\text{0}\leq q \leq \text{1}$ (springpot element) or an element in-between a dashpot ($K_{\text{1}}=\eta$) and an inerter ($K_{\text{2}}=m_R$) when $\text{1}\leq q \leq \text{2}$. In high-frequency microrheology the distributed inertance is $m_R=$ {\large $\frac{m}{\text{6}\pi R}$} with units $\left[ \text{M} \right]\left[ \text{L} \right]^{-\text{1}}$, where $m$ and $R$ are the mass and radius of the particle (bead), respectively \citep{MasonWeitz1995, IndeiSchieberCordoba2012, DominguezGarciaCardinauxBertsevaForroScheffoldJeney2014}.}
\label{fig:Fig02}
\end{figure*}
\begin{equation}\label{eq:Eq37}
M(t)=\frac{K_{\text{2}}}{\Gamma(-\text{2})}\frac{\text{1}}{t^{\text{3}}}=m_R \frac{\mathrm{d}^{\text{2}}\delta(t-\text{0})}{\mathrm{d}t^{\text{2}}}
\end{equation}
which is the memory function of the inerter \citep{Makris2017} and $K_{\text{2}}=m_R$ is its distributed inertance. The result offered by Eq. \eqref{eq:Eq37} is in agreement with the \citet{GelfandShilov1964} alternative definition of the Dirac delta function and its integer-order derivatives offered by Eq. \eqref{eq:Eq31}. Figure \ref{fig:Fig02} shows schematically the Scott-Blair element that is in-between a spring ($K_{\text{0}}=G$) and a dashpot ($K_{\text{1}}=\eta$) when $\text{0}\leq q \leq \text{1}$ or in-between a dashpot ($K_{\text{1}}=\eta$) and an inerter ($K_{\text{2}}=m_R$) when $\text{1}\leq q \leq \text{2}$. \citet{GlockleNonnenmacher1991, GlockleNonnenmacher1993, GlockleNonnenmacher1994} studied the relaxation behavior of fractional derivative rheological models by implementing the Fox H-function and presented plots of the relaxation modulus of the spring -- Scott-Blair element by allowing the fractional derivative of the Scott-Blair element to reach values up to 2 $(\text{0} \leq q \leq \text{2})$, a decade before \citet{Smith2002} introduced the inerter and its direct equivalence to the electric capacitor. In this way \citet{GlockleNonnenmacher1994} have shown plots of the relaxation modulus of what we now call the inerto-elastic fluid \citep{Makris2017}. 

Eqs. \eqref{eq:Eq05} and \eqref{eq:Eq07} indicate that the complex dynamic modulus $\mathcal{G}(\omega)=G_{\text{1}}(\omega)+\operatorname{i}G_{\text{2}}(\omega)$ of the Scott-Blair element given bt Eq. \eqref{eq:Eq26} is
\begin{equation}\label{eq:Eq38}
\mathcal{G}(\omega)=\frac{\tau(\omega)}{\gamma(\omega )}=K_q(\operatorname{i}\omega)^q
\end{equation}
and its inverse Fourier transform is the memory function, $M(t)$, as indicated by Eq. \eqref{eq:Eq10}. With the introduction of the fractional derivative of the Dirac delta function expressed by Eq. \eqref{eq:Eq29} or \eqref{eq:Eq36}, the definition of the memory function given by Eq. \eqref{eq:Eq10} offers a new (to the best of our knowledge) and most useful result regarding the Fourier transform of the function $\mathcal{F}(\omega)=(\operatorname{i}\omega)^q$ with $q \in \mathbb{R}^+$
\begin{align}\label{eq:Eq39}
\mathcal{F}^{-\text{1}}(\operatorname{i}\omega)^q=&\frac{\text{1}}{\text{2}\pi}\int_{-\infty}^{\infty}(\operatorname{i}\omega)^q e^{\operatorname{i}\omega t} \mathrm{d}\omega= \\ \nonumber
&\frac{\mathrm{d}^q\delta(t-\text{0})}{\mathrm{d}t^q}=\frac{\text{1}}{\Gamma(-q)}\frac{\text{1}}{t^{q+\text{1}}} \text{,} \enskip q \in \mathbb{R}^+ \text{,} \enskip t>\text{0}
\end{align}
In terms of the Laplace variable $s=\operatorname{i}\omega$ (see equivalence of Eqs. \eqref{eq:Eq24} and \eqref{eq:Eq25}), Eq. \eqref{eq:Eq39} gives that
\begin{equation}\label{eq:Eq40}
\mathcal{L}^{-\text{1}}\left\lbrace s^q \right\rbrace = \frac{\mathrm{d}^q\delta(t-\text{0})}{\mathrm{d}t^q}=\frac{\text{1}}{\Gamma(-q)}\frac{\text{1}}{t^{q+\text{1}}}  \text{,} \enskip q \in \mathbb{R}^+ \text{,} \enskip t>\text{0}
\end{equation}
where $\mathcal{L}^{-\text{1}}$ 
indicates the inverse Laplace transform operator \citep{LePage1961, Mainardi2010}. While the right-hand side of Eq. \eqref{eq:Eq39} or \eqref{eq:Eq40} is non-zero only when $q\in\left\lbrace\mathbb{R}^+-\mathbb{N}\right\rbrace$ given the poles of the Gamma function when $q$ is zero or any positive integer; and assuming that we are not aware of the \citet{GelfandShilov1964} definition of the Dirac delta function and its integer order derivatives given by Eq. \eqref{eq:Eq31}; the validity of Eq. \eqref{eq:Eq39} can be confirmed by investigating its limiting cases. For instance, when, $q=\text{0}$, $(\operatorname{i}\omega)^q=\text{1}$; and Eq. \eqref{eq:Eq39} yields that $\frac{\text{1}}{\text{2}\pi}\int_{-\infty}^{\infty} e^{\operatorname{i}\omega t} \mathrm{d}\omega=\delta(t-\text{0})$; which is the correct result. When $q=\text{1}$, Eq. \eqref{eq:Eq39} yields that 
\begin{equation}\label{eq:Eq41}
\frac{\text{1}}{\text{2}\pi}\int_{-\infty}^{\infty}\operatorname{i}\omega e^{\operatorname{i}\omega t} \mathrm{d}\omega=\frac{\mathrm{d}\delta(t-\text{0})}{\mathrm{d}t}
\end{equation}
Clearly, the function $\mathcal{F}(\omega)=\operatorname{i}\omega$ is not Fourier integrable in the classical sense, yet the result of Eq. \eqref{eq:Eq41} can be confirmed by evaluating the Fourier transform of {\large $\frac{\mathrm{d}\delta(t-\text{0})}{\mathrm{d}t}$} together with the properties of the higher-order derivatives of the Dirac delta function \citep{Lighthill1958}
\begin{equation}\label{eq:Eq42}
\int_{-\infty}^{\infty}\frac{\mathrm{d}^n\delta(t-\text{0})}{\mathrm{d}t^n}f(t)\mathrm{d}t=(-\text{1})^n\frac{\mathrm{d}^nf(\text{0})}{\mathrm{d}t^n}
\end{equation}
By virtue of Eq. \eqref{eq:Eq42}, the Fourier transform of {\large $\frac{\mathrm{d}\delta(t-\text{0})}{\mathrm{d}t}$} is
\begin{equation}\label{eq:Eq43}
\int_{-\infty}^{\infty} \frac{\mathrm{d}\delta(t-\text{0})}{\mathrm{d}t} e^{-\operatorname{i}\omega t} \mathrm{d}t = -(-\operatorname{i}\omega) e^{-\operatorname{i}\omega \text{0}}=\operatorname{i}\omega
\end{equation}
therefore, the functions $\operatorname{i}\omega$ and {\large $\frac{\mathrm{d}\delta(t-\text{0})}{\mathrm{d}t}$} are Fourier pairs, as indicated by Eq. \eqref{eq:Eq39}.

More generally, for any $q=n \in \mathbb{N}$, Eq. \eqref{eq:Eq39} yields that 
\begin{equation}\label{eq:Eq44}
\frac{\text{1}}{\text{2}\pi}\int_{-\infty}^{\infty} (\operatorname{i}\omega)^n e^{\operatorname{i}\omega t} \mathrm{d}\omega=\frac{\mathrm{d}^n\delta(t-\text{0})}{\mathrm{d}t^n}
\end{equation}
By virtue of Eq. \eqref{eq:Eq42}, the Fourier transform of  {\large $\frac{\mathrm{d}^n\delta(t-\text{0})}{\mathrm{d}t^n}$} is
\begin{equation}\label{eq:Eq45}
\int_{-\infty}^{\infty}\frac{\mathrm{d}^n\delta(t-\text{0})}{\mathrm{d}t^n} e^{\operatorname{i}\omega t} \mathrm{d}t = (-\text{1})^n(-\operatorname{i}\omega)^n = (\operatorname{i}\omega)^n 
\end{equation}
showing that the functions $(\operatorname{i}\omega)^n$ and {\large $\frac{\mathrm{d}^n\delta(t-\text{0})}{\mathrm{d}t^n}$} are \linebreak Fourier pairs, which is a special result (for $q \in \mathbb{N}_{\text{0}}$) of the more general result offered by Eq. \eqref{eq:Eq39}. Consequently, fractional calculus and the memory function of the Scott-Blair element offer an alternative avenue to reach the \citet{GelfandShilov1964} definition of the Dirac delta function and its integer order derivatives given by Eq. \eqref{eq:Eq31}.

The complex dynamic compliance, $\mathcal{J}(\omega)$, of the Scott-Blair element as defined by Eq. \eqref{eq:Eq11} is the inverse of the complex dynamic modulus given by Eq. \eqref{eq:Eq38}
\begin{equation}\label{eq:Eq46}
\mathcal{J}(\omega)=\frac{\gamma(\omega)}{\tau(\omega)}=\frac{\text{1}}{K_q}\frac{\text{1}}{(\operatorname{i}\omega)^q}
\end{equation}
In terms of the Laplace variable, $s=\operatorname{i}\omega$, the impulse fluidity (impulse response function), $\phi (t)$, of the Scott-Blair element is given by
\begin{align}\label{eq:Eq47}
\phi (t) = & \mathcal{L}^{-\text{1}}\left\lbrace\mathcal{J}(s)\right\rbrace =  \\ \nonumber
&\mathcal{L}^{-\text{1}} \left\lbrace \frac{\text{1}}{K_q} \frac{\text{1}}{s^q} \right\rbrace =\frac{\text{1}}{K_q} \frac{\text{1}}{\Gamma (q)} \frac{\text{1}}{t^{\text{1}-q}} U(t-\text{0}) \text{,} \enskip q \in \mathbb{R}^+
\end{align}
The expression for the impulse fluidity (impulse response function) of the Scott-Blair element given by Eq. \eqref{eq:Eq47} has been also presented by \citet{LorenzoHartley2002}. At the limit case where $q=\text{1}$, Eq. \eqref{eq:Eq47} gives $\phi (t) =$ {\large $\frac{\text{1}}{K_{\text{1}}} \frac{\text{1}}{\Gamma (\text{1})} \frac{\text{1}}{t^{\text{0}}}$}$U(t-\text{0})=$ {\large $\frac{\text{1}}{K_{\text{1}}}$}$U(t-\text{0})$, which is the impulse fluidity of the Newtonian fluid with viscosity $\eta = K_{\text{1}}$ (see Table \ref{tab:Table2}). When $q=\text{2}$, Eq. \eqref{eq:Eq47} gives $\phi (t) =$ {\large $\frac{\text{1}}{K_{\text{2}}} \frac{\text{1}}{\Gamma (\text{2})} $}$tU(t-\text{0})=$ {\large $\frac{\text{1}}{K_{\text{2}}}$}$tU(t-\text{0})$, which is the impulse fluidity of the inerter with inertance $m_R = K_{\text{2}}$ \citep{Makris2017, Makris2018}.

The complex dynamic viscosity (impedance), $\eta (\omega)$, of the Scott-Blair element as defined by Eq. \eqref{eq:Eq15} derives directly from Eq. \eqref{eq:Eq26} by using that $\dt{\gamma}(s)=s\gamma(s)$ with $s=\operatorname{i}\omega$. Accordingly, in the Laplace domain, the Scott-Blair element given by Eq. \eqref{eq:Eq26} is expressed as $\tau (s) = K_q s^{q-\text{1}}\dt{\gamma}(s)$ and therefore, the complex dynamic viscosity, $\eta (s)$, of the Scott-Blair element is
\begin{equation}\label{eq:Eq48}
\eta (s) = K_q \frac{\text{1}}{s^{\text{1}-q}} \text{,} \enskip q \in \mathbb{R}^+
\end{equation}

\begin{sidewaystable*}
\centering
\caption{Frequency-response functions and the corresponding causal time-response functions of the classical elementary rheological models and of the Scott-Blair element.}
\setlength{\tabcolsep}{2pt}
{\renewcommand{\arraystretch}{1.5}
\begin{tabularx}{\textheight}{>{\centering}p{0.13\textheight} >{\centering}p{0.165\textheight} >{\centering}p{0.165\textheight} >{\centering}p{0.165\textheight} >{\centering}p{0.165\textheight} >{\centering}p{0.165\textheight} }
	\hline \hline
	& \textbf{Hookean Solid} & \textbf{Newtonian Fluid} & \textbf{Kelvin-Voigt Solid} & \textbf{Maxwell Fluid} & \textbf{Scott-Blair element} 
	\tabularnewline 
	& \includegraphics[scale=0.25]{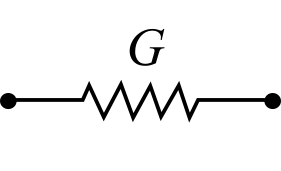} & \includegraphics[scale=0.25]{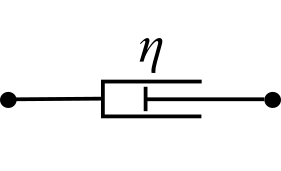} & \includegraphics[scale=0.25]{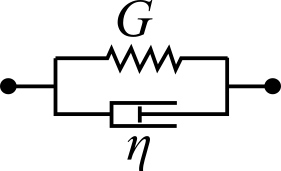} & \includegraphics[scale=0.25]{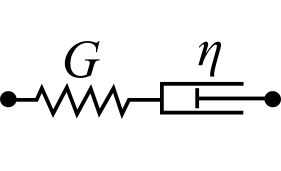}  &  \includegraphics[scale=0.25]{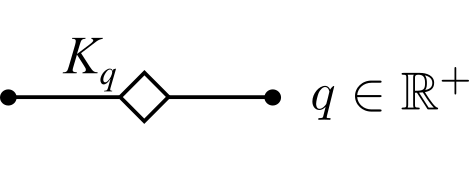}
	\tabularnewline 
	\thead{Constitutive \\ Equation}  & $\tau(t)=G\gamma(t)$ & $\tau(t)=\eta\frac{\mathrm{d}\gamma(t)}{\mathrm{d}t}$& $\tau(t)=G\gamma(t)+\eta\frac{\mathrm{d}\gamma(t)}{\mathrm{d}t}$ & $\tau(t) + \lambda\frac{\mathrm{d}\tau(t)}{\mathrm{d}t}=\eta\frac{\mathrm{d}\gamma(t)}{\mathrm{d}t}$ & $\tau(t)=K_q \frac{\mathrm{d}^q\gamma(t)}{\mathrm{d}t^q}$
	\tabularnewline \hline 
	\thead{Complex \\ Dynamic Modulus \\ $\mathcal{G}(\omega)=\frac{\tau(\omega)}{\gamma(\omega)}$} & {\normalsize $G+\operatorname{i}\text{0}$} & $\text{0}+\operatorname{i}\omega\eta$ & $G+\operatorname{i}\omega\eta$ & $\eta\left[ \cfrac{\lambda\omega^{\text{2}}}{\text{1}+\lambda^{\text{2}}\omega^{\text{2}}} + \operatorname{i}\cfrac{\omega}{\text{1}+\lambda^{\text{2}}\omega^{\text{2}}}  \right]$ & {\normalsize $K_q (\operatorname{i} \omega)^q$}
	\tabularnewline 
	\thead{Complex\\ Dynamic Viscosity \\ $\mathcal{\eta}(\omega)=\frac{\tau(\omega)}{\dt{\gamma}(\omega)}$} & $G\left[\pi  \delta(\omega-\text{0}) -\operatorname{i}\cfrac{\text{1}}{\omega} \right]$ & $\eta+\operatorname{i}\text{0}$ & $G\left[\lambda +\pi  \delta(\omega-\text{0}) -\operatorname{i}\cfrac{\text{1}}{\omega} \right]$ & $\eta \left[ \cfrac{\text{1}}{\text{1}+\lambda^{\text{2}}\omega^{\text{2}}} - \operatorname{i}\cfrac{\omega\lambda}{\text{1}+\lambda^{\text{2}}\omega^{\text{2}}}  \right]$ & $K_q \cfrac{\text{1}}{(\operatorname{i} \omega)^{\text{1}-q}}$
	\tabularnewline  
	\thead{Complex \\ Dynamic Compliance \\ $\mathcal{J}(\omega)=\frac{\text{1}}{\mathcal{G}(\omega)}=\frac{\gamma(\omega)}{\tau(\omega)}$} & $\cfrac{\text{1}}{G}+\operatorname{i}\text{0}$ & $\cfrac{\text{1}}{\eta}\left[ \pi \delta(\omega-\text{0}) -\operatorname{i}\cfrac{\text{1}}{\omega} \right]$ & $\cfrac{\text{1}}{G}\left[ \cfrac{\text{1}}{\text{1}+\lambda^{\text{2}}\omega^{\text{2}}} - \operatorname{i}\cfrac{\omega\lambda}{\text{1}+\lambda^{\text{2}}\omega^{\text{2}}}  \right]$ & $\cfrac{\text{1}}{\eta}\left[ \lambda + \pi\delta(\omega-\text{0}) -\operatorname{i}\cfrac{\text{1}}{\omega} \right]$ & $\cfrac{\text{1}}{K_q} \cfrac{\text{1}}{(\operatorname{i} \omega)^q}$
	\tabularnewline 
	\thead{Complex  \\ Creep Function \\ $\mathcal{C}(\omega)=\frac{\gamma(\omega)}{\dt{\tau}(\omega)}$} & $\cfrac{\text{1}}{G} \left[ \pi\delta(\omega-\text{0}) -\operatorname{i}\cfrac{\text{1}}{\omega} \right]$ & $\cfrac{\text{1}}{\eta}\left[ -\cfrac{\text{1}}{\omega^{\text{2}}} +\operatorname{i}\pi\cfrac{\mathrm{d}\delta(\omega-\text{0})}{\mathrm{d}\omega} \right]$ & \thead{$\frac{\text{1}}{G}\Big[ \pi \delta(\omega-\text{0}) -\operatorname{i}\frac{\text{1}}{\omega}$ \\ \\ $-\frac{\lambda}{\text{1}-\operatorname{i}\omega \lambda}\Big]$} & \thead{$\frac{\text{1}}{G}\left[ \pi \delta(\omega-\text{0}) -\operatorname{i}\frac{\text{1}}{\omega}\right] +$ \\ \\ $+\frac{\text{1}}{\eta}\left[ -\frac{\text{1}}{\omega^{\text{2}}} +\operatorname{i}\pi\frac{\mathrm{d}\delta(\omega-\text{0})}{\mathrm{d}\omega} \right]$} & $\cfrac{\text{1}}{K_q} \cfrac{\text{1}}{(\operatorname{i} \omega)^{q+\text{1}}}$ 	\tabularnewline  
	\thead{ Complex \\ Dynamic Fluidity\\ $\mathcal{\phi}(\omega)=\frac{\text{1}}{\eta(\omega)}=\frac{\dt{\gamma}(\omega)}{\tau(\omega)}$} & $\text{0}+\operatorname{i}\omega\cfrac{\text{1}}{G}$ & $\cfrac{\text{1}}{\eta}+\operatorname{i}\text{0}$ & $\cfrac{\text{1}}{G}\left[ \cfrac{\lambda\omega^{\text{2}}}{\text{1}+\lambda^{\text{2}}\omega^{\text{2}}} + \operatorname{i}\cfrac{\omega}{\text{1}+\lambda^{\text{2}}\omega^{\text{2}}}  \right]$  & $\cfrac{\text{1}}{\eta} \left[ \text{1} + \operatorname{i}\lambda\omega \right]$ & $\cfrac{\text{1}}{K_q} (\operatorname{i} \omega)^{\text{1}-q}$
	\tabularnewline  \hline 
	\thead{Memory Function \\ $M(t)$} & $G\delta(t-\text{0})$ & $\eta\cfrac{\mathrm{d}\delta(t-\text{0})}{\mathrm{d}t}$ & $G \left[ \delta(t-\text{0}) +\lambda \frac{\mathrm{d}\delta(t-\text{0})}{\mathrm{d}t} \right]$ & $G \left[ \delta(t-\text{0}) - \frac{\text{1}}{\lambda} e^{-t/\lambda} \right]$ &  $\cfrac{K_q}{\Gamma(-q)}\cfrac{\text{1}}{t^{q+\text{1}}} U(t-\text{0})$
	\tabularnewline  
	\thead{Relaxation Modulus \\ $G(t)$} & $GU(t-\text{0})$ & $\eta\delta(t-\text{0})$ & $G \left[ \lambda \delta(t-\text{0}) +U (t-\text{0})\right]$  & $G  e^{-t/\lambda}$ & $\cfrac{K_q}{\Gamma(\text{1}-q)}\cfrac{\text{1}}{t^q} U(t-\text{0})$
	\tabularnewline 
	\thead{Impulse Fluidity \\ $\phi(t)$} & $\cfrac{\text{1}}{G} \,\delta(t-\text{0})$ & $\cfrac{\text{1}}{\eta} \, U(t-\text{0})$ & $\cfrac{\text{1}}{\eta} \, e^{-t/\lambda} $ & $\cfrac{\text{1}}{\eta} \left[ \lambda \delta(t-\text{0}) +U(t-\text{0})\right]$ & $\cfrac{\text{1}}{K_q} \cfrac{\text{1}}{\Gamma(q)} \cfrac{\text{1}}{t^{\text{1}-q}} U(t-\text{0})$
	\tabularnewline 
	\thead{Creep Compliance \\ $J(t)$}	& $\cfrac{\text{1}}{G} \, U(t-\text{0})$ & $\cfrac{\text{1}}{\eta} \, t \, U(t-\text{0})$ & $\cfrac{\text{1}}{G} \, \left[U(t-\text{0}) - e^{-t/\lambda} \right]$ & $\cfrac{\text{1}}{G} \left(  \text{1} + \cfrac{t}{\lambda} \right)  U(t-\text{0}) $ & $\cfrac{\text{1}}{K_q} \cfrac{\text{1}}{\Gamma(q+\text{1})} t^q U(t-\text{0})$
	\tabularnewline 
	\thead{Impulse Strain-rate \\ Response Function \\ $\psi(t)$}	& $\cfrac{\text{1}}{G} \cfrac{\mathrm{d}\delta(t-\text{0})}{\mathrm{d}t}$ & $\cfrac{\text{1}}{\eta} \, \delta(t-\text{0})$ & $\cfrac{\text{1}}{\eta} \left[\delta(t-\text{0}) - \cfrac{\text{1}}{\lambda} e^{-t/\lambda} \right]$ & $\cfrac{\text{1}}{\eta} \left[ \delta(t-\text{0}) +\lambda \frac{\mathrm{d}\delta(t-\text{0})}{\mathrm{d}t} \right]$ & $\cfrac{\text{1}}{K_q} \cfrac{\text{1}}{\Gamma(-1+q)} \cfrac{\text{1}}{t^{\text{2}-q}}U(t-\text{0})$
	\tabularnewline \hline \hline
\end{tabularx}}
\label{tab:Table2}
\end{sidewaystable*}

\noindent For the springpot element $(\text{0}\leq q \leq \text{1})$ which is a special case of the Scott-Blair element $(q \in \mathbb{R}^+)$ the quantity $\text{1}-q>\text{0}$, and the relaxation modulus, $G(t)$, of the springpot element is offered by the classical result available in Tables of Laplace transforms \citep{Bateman1954}
\begin{align}\label{eq:Eq49}
G(t)=&\mathcal{L}^{-\text{1}} \left\lbrace \eta(s) \right\rbrace = \mathcal{L}^{-\text{1}} \left\lbrace K_q \frac{\text{1}}{s^{\text{1}-q}} \right\rbrace = \\ \nonumber
&K_q \frac{\text{1}}{\Gamma(\text{1}-q)}\frac{\text{1}}{t^q}U(t-\text{0}) \text{,} \enskip \text{0} < q < \text{1}
\end{align}
The result offered by Eq. \eqref{eq:Eq49} is well known to the literature \citep{SmitDeVries1970, Koeller1984, Friedrich1991, HeymansBauwens1994, SukiBarabasiLutchen1994, SchiesselMetzlerBlumenNonnenmacher1995, PaladeVerneyAttane1996, CraiemMagin2010, Mainardi2010}. For the case where $q>\text{1}$ (say Scott-Blair element that is in between a dashpot and an inerter, $\text{1}\leq q\leq \text{2}$), the Laplace transform offered by the right-hand side of Eq. \eqref{eq:Eq49} does not exist in the classical sense and one has to use the result of Eq. \eqref{eq:Eq40}. Accordingly for $q>\text{1}$, the complex dynamic viscosity of the Scott-Blair element is $\eta (s)=K_q s^{q-\text{1}}$ and Eq. \eqref{eq:Eq40} yields
\begin{align}\label{eq:Eq50}
G(t)= & \mathcal{L}^{-\text{1}} \left\lbrace K_q s^{q-\text{1}} \right\rbrace = \frac{K_q}{\Gamma(-q+\text{1})}\frac{\text{1}}{t^{q-\text{1}+\text{1}}}U(t-\text{0})= \\ \nonumber
&\frac{K_q}{\Gamma(\text{1}-q)}\frac{\text{1}}{t^q}U(t-\text{0})\text{,} \enskip  q > \text{1}
\end{align}
Interestingly, the result offered by Eq. \eqref{eq:Eq50} for $q>1$ is identical to the classical result offered by Eq. \eqref{eq:Eq49} for $\text{0} \leq q < \text{1}$; therefore Eq. \eqref{eq:Eq49} is the expression of the relaxation modulus of the Scott-Blair element for any $q \in \mathbb{R}^+$. At the limit case where $q=\text{0}$, Eq. \eqref{eq:Eq49} gives $G(t)=K_{\text{0}}${\large $\frac{\text{1}}{\Gamma(\text{1})}\frac{\text{1}}{t^{\text{0}}}$}$U(t-\text{0})=K_{\text{0}}U(t-\text{0})$ which is the relaxation modulus of the Hookean spring with elastic modulus $G=K_{\text{0}}$ (see Table \ref{tab:Table2}). When $q=\text{1}$, Eq. \eqref{eq:Eq50} becomes the Dirac delta function, $\delta(t-\text{0})$, according to the definition given by Eq. \eqref{eq:Eq31} \citep{GelfandShilov1964}; therefore, $G(t)=K_{\text{1}}\delta(t-\text{0})$ which is the relaxation modulus of the Newtonian dashpot with viscosity $\eta=K_{\text{1}}$ (see Table \ref{tab:Table2}). When $q=\text{2}$, Eq. \eqref{eq:Eq50} yields $G(t)=$ {\large $\frac{K_{\text{2}}}{\Gamma(-\text{1})}\frac{\text{1}}{t^{\text{2}}}$} $=K_{\text{2}}${\large $\frac{\mathrm{d}\delta(t-\text{0})}{\mathrm{d}t}$} which is the relaxation modulus of the inerter with inertance $m_R=K_{\text{2}}$ \citep{Makris2017, Makris2018}.

The complex dynamic fluidity (admittance), $\phi (\omega)$, of the Scott-Blair element as defined by Eq. \eqref{eq:Eq18} is the inverse of its complex dynamic viscosity given by Eq. \eqref{eq:Eq48}
\begin{equation}\label{eq:Eq51}
\phi (s)=\frac{\text{1}}{K_q} s^{\text{1}-q}\text{,} \enskip \text{0} < q < \text{1}
\end{equation} 
For the special case of the springpot element $(\text{0}\leq q \leq \text{1})$, $\text{1}-q \geq \text{0}$, the impulse strain-rate response function of the springpot element is offered with the help of Eq. \eqref{eq:Eq40}
\begin{align}\label{eq:Eq52}
\psi (t) = &\mathcal{L}^{-\text{1}}\left\lbrace \phi (s) \right\rbrace = \mathcal{L}^{-\text{1}}\left\lbrace \frac{\text{1}}{K_q}s^{\text{1}-q} \right\rbrace = \\ \nonumber
&\frac{\text{1}}{K_q} \frac{\text{1}}{\Gamma (-\text{1}+q)} \frac{\text{1}}{t^{\text{2}-q}}U(t-\text{0}) \text{,} \enskip \text{0} \leq q \leq \text{1}
\end{align}
For the case where $q>\text{1}$ (say the Scott-Blair element that is in between a dashpot and an inerter: $\text{1}<q<\text{2}$) the complex dynamic fluidity of the Scott-Blair element is $\phi (s)=$ {\large $\frac{\text{1}}{K_q} \frac{\text{1}}{s^{q-\text{1}}}$}, and its inverse Laplace transform is offered from the classical result available in Tables of Laplace Transforms \citep{Bateman1954}
\begin{align}\label{eq:Eq53}
\psi (t) = & \mathcal{L}^{-\text{1}}\left\lbrace \phi (s) \right\rbrace = \mathcal{L}^{-\text{1}}\left\lbrace \frac{\text{1}}{K_q}\frac{\text{1}}{s^{q-\text{1}}} \right\rbrace = \\ \nonumber
& \frac{\text{1}}{K_q} \frac{\text{1}}{\Gamma (-\text{1}+q)} \frac{\text{1}}{t^{\text{2}-q}}U(t-\text{0}) \text{,} \enskip q > \text{1}
\end{align}
The classical result offered by Eq. \eqref{eq:Eq53} for $q > \text{1}$ is identical to the result of Eq. \eqref{eq:Eq52} for $\text{0} < q < \text{1}$; therefore Eq. \eqref{eq:Eq53} is the expression of the impulse strain-rate response function of the Scott-Blair element for any $q \in \mathbb{R}^+$. For the limit cases where $q=\text{0}$ and $K_q=G$ or $q=\text{1}$ and $K_q=\eta$, Eq. \eqref{eq:Eq53} results by virtue of Eq. \eqref{eq:Eq31} that $\psi(t)=$ {\large $\frac{\text{1}}{G}\frac{\mathrm{d}\delta(t-\text{0})}{\mathrm{d}t}$} or $\psi(t)=$ {\large $\frac{\text{1}}{\eta}$}$\delta(t-\text{0})$ which are respectively the impulse strain-rate response functions of the Hookean spring or the Newtonian dashpot, as shown in Table \ref{tab:Table2}. For the limit case where $q=\text{2}$, Eq. \eqref{eq:Eq53} results, $\psi(t)=$ {\large $\frac{\text{1}}{K_{\text{2}}}$}$U(t-\text{0})$ which is the impulse strain-rate response function of the inerter with inertance $m_R=K_{\text{2}}$ \citep{Makris2017, Makris2018}.

The complex creep function, $\mathcal{C}(\omega)$, of the Scott-Blair element as defined by Eq. \eqref{eq:Eq22} derives directly from equation \eqref{eq:Eq46} given that $\mathcal{C}(s)=$ {\large $\frac{\mathcal{J}(s)}{s}$} with $s=\operatorname{i}\omega$. Accordingly, the complex creep function, $\mathcal{C}(s)$, of the Scott-Blair element is
\begin{equation}\label{eq:Eq54}
\mathcal{C}(s)=\frac{\text{1}}{K_q} \frac{\text{1}}{s^{q+\text{1}}} \text{,} \enskip q \in \mathbb{R}^+
\end{equation}
Given that $q+\text{1}>\text{0}$, the creep compliance of the Scott-Blair element is offered by the classical results available in Tables of Laplace transforms \citep{Bateman1954}
\begin{align}\label{eq:Eq55}
J(t)=&\mathcal{L}^{-\text{1}}\left\lbrace \mathcal{C}(s) \right\rbrace = \mathcal{L}^{-\text{1}}\left\lbrace \frac{\text{1}}{K_q} \frac{\text{1}}{s^{q+\text{1}}} \right\rbrace = \\ \nonumber
& \frac{\text{1}}{K_q} \frac{\text{1}}{\Gamma(q+\text{1})} t^q U(t-\text{0}) \text{,} \enskip q \in \mathbb{R}^+
\end{align}
The expression given by Eq. \eqref{eq:Eq55} has been presented by \citet{Koeller1984, Friedrich1991, HeymansBauwens1994, SchiesselMetzlerBlumenNonnenmacher1995}. For the limit cases when $q=\text{0}$ and $K_q=G$ or $q=\text{1}$ and $K_q=\eta$, Eq. \eqref{eq:Eq55} results that $J(t)=$ {\large $\frac{\text{1}}{G}$}$U(t-\text{0})$ or $J(t)=$ {\large $\frac{\text{1}}{\eta}$}$tU(t-\text{0})$ which are respectively the creep compliances of the Hookean spring or the Newtonian dashpot shown in Table \ref{tab:Table2}.

The five causal time-response functions of the Scott-Blair (springpot) element computed in this section (Eqs. \eqref{eq:Eq36}, \eqref{eq:Eq47}, \eqref{eq:Eq49}, \eqref{eq:Eq52} and \eqref{eq:Eq55}) are summarized in Table \ref{tab:Table2} next to the known time-response function of the Hookean spring, the Newtonian dashpot, the Kelvin-Voigt solid and the Maxwell fluid \citep{HarrisCrede1976, BirdArmstrongHassager1987, Giesekus1995, Makris1997b} which are included to validate the limit-cases of the results derived for the generalized fractional derivative rheological models examined in this work.

\begin{figure*}[htp]
\centering
\includegraphics[width=.72\textwidth, angle=0]{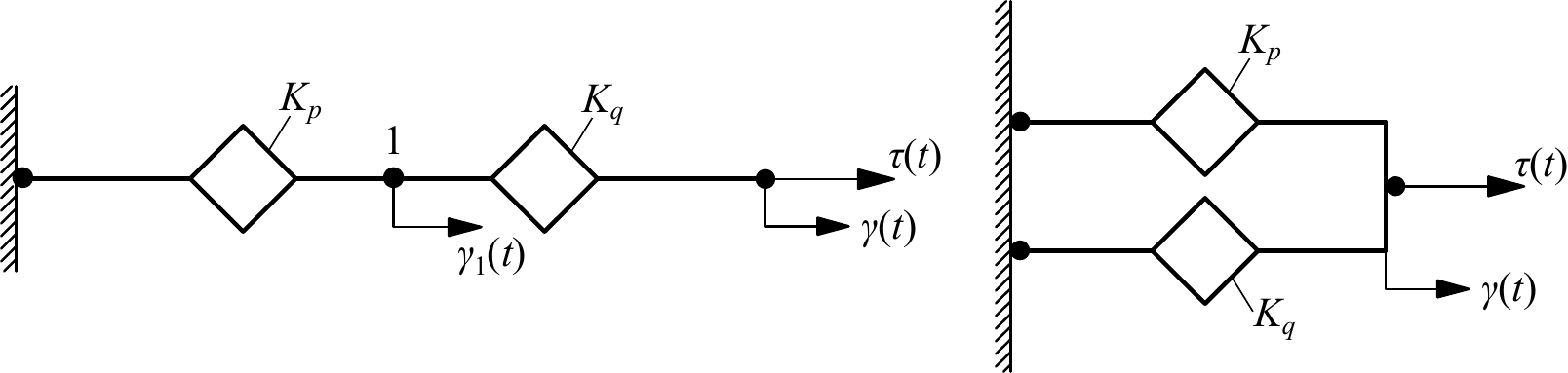}
\caption{The generalized fractional Maxwell fluid (left) and the generalized fractional Kelvin-Voigt element (right) with $p$, $q \in \mathbb{R}^+$.}
\label{fig:Fig03}
\end{figure*}

\section{Two-parameter fractional derivative rheological models}
Upon we introduced the fractional derivative of the Dirac delta function and derived the causal time-re\-sponse functions of the elementary Scott-Blair element expressed by Eq. \eqref{eq:Eq26}; this study proceeds with the derivation of the time-response functions of the two-parameter generalized fractional Maxwell fluid and generalized fractional Kelvin-Voigt element. The generalized fractional Maxwell fluid consists of two Scott-Blair elements connected in series as shown in Figure \ref{fig:Fig03} (left); while the generalized fractional Kelvin-Voigt element consists of two Scott-Blair elements connected in parallel as shown in Figure \ref{fig:Fig03} (right).

Because of the linearity of the Scott-Blair element, the basic response functions of the fractional rheological models showed in Figure \ref{fig:Fig03} follow the same superposition rules that govern the basic response functions of classical linear networks. For instance, the complex dynamic compliance (dynamic flexibility), complex dynamic fluidity (admittance) and complex creep function (any transfer function that has a strain or a strain-rate on its numerator) of the generalized fractional Maxwell fluid (in-series connection) are the summation of the corresponding dynamic compliances, dynamic fluidities or complex creep functions of the individual Scott-Blair elements. The outcome of this superposition is reflected in the resulting causal time-response functions which are the impulse fluidity (impulse response function), impulse strain-rate function or creep compliance (retardation function).

Similarly, the complex dynamic modulus (dynamic stiffness) and complex dynamic viscosity (impedance -- that is any transfer function that has a stress on its numerator) of the generalized fractional Kelvin-Voigt element (in-parallel connection) are the summation of the corresponding dynamic moduli or dynamic viscosities of the individual Scott-Blair elements. The outcome of this superposition is reflected in the resulting causal time-response functions which are the memory function or the relaxation modulus.

A function that is central in the derivation of the time-response functions of the fractional-derivative rheological models examined in this study is the two-para\-meter Mittag-Leffler function \citep{Bateman1953, Podlubny1998, HauboldMathaiSaxena2011, GorenfloKilbasMainardiRogosin2014}
\begin{equation}\label{eq:Eq56}
E_{\alpha\text{, }\beta}(z)=\sum\limits_{j=\text{0}}^{\infty}\frac{z^j}{\Gamma(j\alpha+\beta)} \text{,} \enskip \alpha > \text{0} \text{, } \beta > \text{0}  
\end{equation}
The evaluation of some time-response functions of the generalized fractional derivative rheological models examined in this paper involves the fractional derivative of the Mittag-Leffler function and this may result to negative values of $\beta$ $(\beta<\text{0} \text{, see Eq. \eqref{eq:Eq59}})$. When this happens, the singularities embedded in the resulting Mittag-Leffler function are extracted using the recurrence relation \citep{Bateman1953, HauboldMathaiSaxena2011}
\begin{equation}\label{eq:Eq57}
E_{\alpha\text{, }\beta}(z)=\frac{\text{1}}{\Gamma(\beta)}+zE_{\alpha\text{, }\alpha+\beta}(z)
\end{equation}
Of interset in this paper are the fractional integral and the fractional derivative of the function \break $\mathit{\varUpsilon}_{\alpha\text{, }\beta}(t)= t^{\beta-\text{1}}E_{\alpha\text{, }\beta}${\large $\Big($$-$$\frac{\text{1}}{\lambda}$}$t^{\alpha}${\large $\Big)$}, which is the product of a power law with the Mittag-Leffler function. When $\alpha=\beta$, this function is also known as the Rabotnov function \citep{Rabotnov1980}. The fractional integral of $\mathit{\varUpsilon}_{\alpha\text{, }\beta}(t)$ is
\begin{align}\label{eq:Eq58}
&\frac{\text{1}}{\Gamma(q)}\int_{\text{0}}^{t} (t-\xi)^{q-\text{1}} \xi^{\beta-\text{1}} E_{\alpha\text{, }\beta}\left(- \frac{\text{1}}{\lambda} \xi^{\alpha} \right) \mathrm{d}\xi = \\ \nonumber
& I^q\left[ t^{\beta-\text{1}} E_{\alpha\text{, }\beta}\left( -\frac{\text{1}}{\lambda} t^{\alpha} \right) \right] = t^{\beta+q-\text{1}}  E_{\alpha\text{, }\beta+q}\left(- \frac{\text{1}}{\lambda} t^{\alpha} \right)
\end{align}
while its fractional derivative is
\begin{equation}\label{eq:Eq59}
\frac{\mathrm{d}^q}{\mathrm{d}t^q} \left[ t^{\beta-\text{1}}  E_{\alpha\text{, }\beta}\left(- \frac{\text{1}}{\lambda} t^{\alpha} \right) \right] =  t^{\beta-q-\text{1}} E_{\alpha\text{, }\beta-q}\left(- \frac{\text{1}}{\lambda} t^{\alpha} \right)
\end{equation}
In the event that $\beta-q < $ 0, the Mittag-Leffler function appearing on the right-hand side of Eq. \eqref{eq:Eq59} is replaced with the identity from the recurrence relation \eqref{eq:Eq57}
\begin{align}\label{eq:Eq60}
&\frac{\mathrm{d}^q}{\mathrm{d}t^q}  \left[ t^{\beta-\text{1}}  E_{\alpha\text{, }\beta}\left(- \frac{\text{1}}{\lambda} t^{\alpha} \right) \right] = \\ \nonumber
&\frac{\text{1}}{\Gamma(\beta-q)} \frac{\text{1}}{t^{\text{1}+q-\beta}} - \frac{\text{1}}{\lambda} t^{\alpha+\beta-q-\text{1}}  E_{\alpha\text{, }\alpha+\beta-q}\left(- \frac{\text{1}}{\lambda} t^{\alpha} \right)
\end{align}
Recognizing that according to Eq. \eqref{eq:Eq30} the first term in the right-hand side of Eq. \eqref{eq:Eq60} is {\large $\frac{\mathrm{d}^{q-\beta}}{\mathrm{d}t^{q-\beta}}$}$\delta(t-\text{0})$, Eq. \eqref{eq:Eq60} is expressed as
\begin{align}\label{eq:Eq61}
&\frac{\mathrm{d}^q}{\mathrm{d}t^q}  \left[ t^{\beta-\text{1}}  E_{\alpha\text{, }\beta}\left(- \frac{\text{1}}{\lambda} t^{\alpha} \right) \right] = \\ \nonumber
&\frac{\mathrm{d}^{q-\beta}}{\mathrm{d}t^{q-\beta}}\delta(t-\text{0}) - \frac{\text{1}}{\lambda} t^{\alpha+\beta-q-\text{1}}  E_{\alpha\text{, }\alpha+\beta-q}\left(- \frac{\text{1}}{\lambda} t^{\alpha} \right)
\end{align}
where the singularity {\large $\frac{\mathrm{d}^{q-\beta}}{\mathrm{d}t^{q-\beta}}$}$\delta(t-\text{0})$ has been extracted from the right-hand side of Eq. \eqref{eq:Eq59} and now, the second index of the Mittag-Leffler function has been increased to $\alpha+\beta-q$. In the event that $\alpha+\beta-q$ remains negative $(\alpha+\beta-q< \text{0})$, the Mittag-Leffler function appearing on the right-hand side of Eq. \eqref{eq:Eq60} or \eqref{eq:Eq61} is replaced again by virtue of the recurrence relation \eqref{eq:Eq57} until all singularities are extracted.

\section{Time-response functions of the generalized fractional Maxwell fluid}
With reference to Figure \ref{fig:Fig03} (left), the stress, $\tau(t)$ \linebreak (through variable) is common in both Scott-Blair elements that are connected in series. With this configuration
\begin{equation}\label{eq:Eq62}
\tau(t)=K_p\frac{\mathrm{d}^p\gamma_{\text{1}}(t)}{\mathrm{d}t^p} \text{,} \enskip p \in \mathbb{R}^+
\end{equation}
and at the same time
\begin{equation}\label{eq:Eq63}
\tau(t)=K_q\frac{\mathrm{d}^q\left(\gamma(t)-\gamma_{\text{1}}(t)\right)}{\mathrm{d}t^q} \text{,} \enskip q \in \mathbb{R}^+
\end{equation}
where $\gamma_{\text{1}}(t)=$ nodal displacement of the internal node 1. Without loss of generality we assume $p<q$ and we take the $q-p>\text{0}$ fractional derivative of Eq. \eqref{eq:Eq62}
\begin{equation}\label{eq:Eq64}
\frac{\mathrm{d}^{q-p}}{\mathrm{d}t^{q-p}} \tau(t)=K_q \frac{\mathrm{d}^{q-p}}{\mathrm{d}t^{q-p}} \frac{\mathrm{d}^p\gamma_{\text{1}}(t)}{\mathrm{d}t^p}=K_p\frac{\mathrm{d}^q\gamma_{\text{1}}(t)}{\mathrm{d}t^q}
\end{equation}
Substitution of {\large $\frac{\mathrm{d}^q\gamma_{\text{1}}(t)}{\mathrm{d}t^q}$} given by Eq. \eqref{eq:Eq64} into Eq. \eqref{eq:Eq63} gives
\begin{equation}\label{eq:Eq65}
\tau(t)+\frac{K_q}{K_p} \frac{\mathrm{d}^{q-p}}{\mathrm{d}t^{q-p}}\tau(t)=K_q \frac{\mathrm{d}^q\gamma(t)}{\mathrm{d}t^q}
\end{equation}
Eq. \eqref{eq:Eq65} has been presented by \citet{Friedrich1991, SchiesselMetzlerBlumenNonnenmacher1995} and was used by \citet{JaishankarMcKinley2013} to describe the interfacial rheological properties between bovine serum albumin (BSA) and Acacia gum solutions. When $q=\text{1}$, the generalized fractional Maxwell model given by Eq. \eqref{eq:Eq65} reduces to a springpot -- dashpot in-series connection --- a model that was proposed by \citet{Makris1992, MakrisConstantinou1991, MakrisConstantinou1992} to describe the behavior of viscoelastic fluid dampers that find applications in vibration and seismic isolation \citep{MakrisDeoskar1996}. When $p=\text{0}$, the fractional Maxwell model given by Eq. \eqref{eq:Eq65} reduces to a spring -- Scott-Blair in-series connection \citep{MainardiSpada2011}. 

By using $r=q-p>\text{0}$ and $\lambda_r=$ {\large $\nicefrac{K_q}{K_p}$}, the Fourier transform of Eq. \eqref{eq:Eq65} gives
\begin{equation}\label{eq:Eq66}
\tau(\omega)\left[ \text{1}+\lambda_r (\operatorname{i}\omega)^r \right] = K_q (\operatorname{i}\omega)^q \gamma (\omega)
\end{equation}
and the dynamic modulus, $\mathcal{G}(s)$ with $s=\operatorname{i}\omega$ of the generalized fractional Maxwell fluid given by Eq. \eqref{eq:Eq65} is
\begin{equation}\label{eq:Eq67}
\mathcal{G}(s)=K_q \frac{s^q}{\text{1}+\lambda_rs^r}=K_p\frac{s^q}{s^r+\nicefrac{K_p}{K_q}}
\end{equation}
The inverse Laplace transform of Eq. \eqref{eq:Eq67} is evaluated with the convolution integral \citep{LePage1961}
\begin{align}\label{eq:Eq68}
M(t)=&\mathcal{L}^{-\text{1}}\left\lbrace \mathcal{G}(s) \right\rbrace = \\ \nonumber
&\mathcal{L}^{-\text{1}}\left\lbrace \mathcal{F}(s)\mathcal{H}(s) \right\rbrace = \int_{\text{0}}^t f(t-\xi)h(\xi) \mathrm{d}\xi
\end{align}
where according to Eq. \eqref{eq:Eq40}, 
\begin{align}\label{eq:Eq69}
f(t)=&\mathcal{L}^{-\text{1}}\left\lbrace \mathcal{F}(s) \right\rbrace = \\ \nonumber
&\mathcal{L}^{-\text{1}}\left\lbrace K_ps^q \right\rbrace = \frac{K_p}{\Gamma(-q)}\frac{\text{1}}{t^{q+\text{1}}}  \text{,} \enskip q \in \mathbb{R}^+
\end{align}
and
\begin{align}\label{eq:Eq70}
h(t)=&\mathcal{L}^{-\text{1}}\left\lbrace \mathcal{H}(s) \right\rbrace = \mathcal{L}^{-\text{1}}\left\lbrace \frac{\text{1}}{s^r+\nicefrac{K_p}{K_q}}\right\rbrace =\\ \nonumber
& t^{r-\text{1}} E_{r\text{, }r}\left( -\frac{K_p}{K_q}t^r \right)  \text{,} \enskip r<q \in \mathbb{R}^+
\end{align}
where $E_{\alpha\text{, }\beta}(z)$ is the Mittag-Leffler function defined by Eq. \eqref{eq:Eq56}. Substitution of the results of Eqs. \eqref{eq:Eq69} and \eqref{eq:Eq70} into the convolution given by Eq. \eqref{eq:Eq68}, the memory function of the generalized fractional derivative Maxwell fluid is
\begin{align}\label{eq:Eq71}
M(t)=&\mathcal{L}^{-\text{1}}\left\lbrace \mathcal{G}(s) \right\rbrace= \\ \nonumber
&\frac{K_p}{\Gamma(-q)}\int_{\text{0}}^t \frac{\text{1}}{(t-\xi)^{q+\text{1}}}\frac{\text{1}}{\xi^{\text{1}-r}}E_{r\text{, }r}\left( -\frac{K_p}{K_q}\xi^r \right)  \mathrm{d}\xi
\end{align}
Eq. \eqref{eq:Eq71} shows that the memory function, $M(t)$, of the generalized fractional Maxwell model is merely the fractional derivative of order $q$ (see Eq. \eqref{eq:Eq04}) of the function given by Eq. \eqref{eq:Eq70} $(r=q-p)$
\begin{align}\label{eq:Eq72}
M(t)=&K_p\frac{\mathrm{d}^q}{\mathrm{d}t^q}\left[ \frac{\text{1}}{t^{\text{1}-q+p}}E_{q-p\text{, }q-p}\left( -\frac{K_p}{K_q}t^{q-p} \right)  \right] = \\ \nonumber
&K_p \frac{\text{1}}{t^{\text{1}+p}} E_{q-p\text{, }-p} \left( -\frac{K_p}{K_q} t^{q-p}\right)
\end{align}
where the right-hand side of Eq. \eqref{eq:Eq72} was obtained by using the result of Eq. \eqref{eq:Eq59} with $\beta = q-p$.

Given that the second index of the Mittag-Leffler function appearing in the right-hand side of Eq. \eqref{eq:Eq72} is negative $(-p<\text{0})$, the singularity embedded in the memory function, $M(t)$, of the generalized fractional Maxwell model is extracted by virtue of Eq. \eqref{eq:Eq61} with $\beta = q-p$
\begin{align}\label{eq:Eq73}
M(t)= & K_p \Bigg[ \frac{\mathrm{d}^p}{\mathrm{d}t^p} \delta(t-\text{0})- \\ \nonumber
& \frac{K_p}{K_q} \frac{\text{1}}{t^{\text{1}+\text{2}p-q}} E_{q-p\text{, }q-\text{2}p}\left( -\frac{K_p}{K_q} t^{q-p} \right)  \Bigg]
\end{align}
In the event that $q-\text{2}p$ remains negative $(q-\text{2}p<\text{0})$, application once again of the recurrence relationship given by Eq. \eqref{eq:Eq57} to the MIttag-Leffler function appearing in the right-hand side of Eq. \eqref{eq:Eq73} gives
\begin{align}\label{eq:Eq74}
M(t)=& K_p \Bigg[\frac{\mathrm{d}^p}{\mathrm{d}t^p} \delta(t-\text{0}) - \frac{K_p}{K_q} \frac{\mathrm{d}^{\text{2}p-q}}{\mathrm{d}t^{\text{2}p-q}} \delta(t-\text{0}) + \\ \nonumber
&\left( \frac{K_p}{K_q}\right)^{\text{2}}\frac{\text{1}}{t^{\text{1}+\text{3}p-\text{2}q}} E_{q-p\text{, }\text{2}q-\text{3}p}\left( -\frac{K_p}{K_q} t^{q-p} \right)  \Bigg]
\end{align}
where now the next singularity {\large $\frac{\mathrm{d}^{\text{2}q-p}}{\mathrm{d}t^{\text{2}q-p}}$}$\delta(t-\text{0})$ has been extracted. In the event that $\text{2}q-\text{3}p$ remains negative, this procedure will be repeated until the second index of the Mittag-Leffler function appearing in the right-hand side of the memory function, $M(t)$, is positive and in this way all singularities will have been extracted.

\noindent \textbf{Special Cases:} 1. \textit{Spring -- Scott-Blair element in-series} $(p=\text{0}$, $r=q-p=q \in \mathbb{R}^+ \text{, } K_p=G)$. In this case where $p=\text{0}$, we use the result for the memory function, $M(t)$, offered by of Eq. \eqref{eq:Eq73}
{\small \begin{equation}\label{eq:Eq75}
M(t) = G\left[ \delta(t-\text{0}) - \frac{G}{K_q} \frac{\text{1}}{q^{\text{1}-q}} E_{q\text{, }q} \left( -\frac{G}{K_q} t^{q} \right) \right] \text{,} \enskip q \in \mathbb{R}^+
\end{equation}}
Alternatively, for this special case where $p=\text{0}$, the singularity $\delta(t-\text{0})$ embedded in $M(t)$ as shown by Eq. \eqref{eq:Eq75} can be extracted by expanding the dynamic modulus $\mathcal{G}(s)$ given by Eq. \eqref{eq:Eq67} $(r=q-p=q \text{ and } K_P=G)$ into partial fractions
\begin{equation}\label{eq:Eq76}
\mathcal{G}(s)=G\left[ \text{1} - \frac{G}{K_q} \frac{\text{1}}{s^q+\nicefrac{G}{K_q}} \right] \text{,} \enskip q \in \mathbb{R}^+
\end{equation}
By virtue of Eq. \eqref{eq:Eq70} in association with that $\mathcal{L}^{-\text{1}}\left\lbrace \text{1} \right\rbrace=\delta(t-\text{0})$, the inverse Laplace transform of Eq. \eqref{eq:Eq76} gives precisely the result of Eq. \eqref{eq:Eq75}. 
When $q=\text{1}$ and $K_q=\eta$, the dynamic modulus given by Eq. \eqref{eq:Eq76} reduces to that of the classical Maxwell model. In this case Eq. \eqref{eq:Eq75} gives
\begin{equation}\label{eq:Eq77}
M(t)=G\left[ \delta(t-\text{0}) -\frac{G}{\eta} E_{\text{1, 1}} \left( -\frac{G}{\eta}t \right)\right]
\end{equation}
Using the identity that $E_{\text{1, 1}}(z)=E_{\text{1}}(z)=e^z$, together with {\large $\nicefrac{\eta}{G}=$} $\lambda=$ relaxation time, Eq. \eqref{eq:Eq77} gives that the memory function of the classical Maxwell model $(p=\text{0, }q=\text{1})$ is $M(t)=G \Big[ \delta(t-\text{0})-${\large $\frac{\text{1}}{\lambda}$}$e${\large $^{-\nicefrac{t}{\lambda}}$}$\Big]$, which is the classical result appearing in Table \ref{tab:Table2}. When $q=\text{2}$, $K_q=K_{\text{2}}=m_R$ $\big($with units $\left[ \text{M} \right]\left[ \text{L} \right]^{-\text{1}}\big)$ is the distributed inertance of an inerter connected in-series with a Hookean spring with elastic constant $G$. In this case Eq. \eqref{eq:Eq75} gives
\begin{equation}\label{eq:Eq78}
M(t)=G\left[ \delta(t-\text{0}) -\frac{G}{m_R} t E_{\text{2, 2}}\left( -\frac{G}{m_R}t^{\text{2}} \right) \right] 
\end{equation}
By using that {\large $\frac{G}{m_R}$} $=\omega_R^{\text{2}}$, where $\omega_R$ is the rotational frequency of a spring -- inerter in-series connection \citep{Makris2017, Makris2018} together with the identity
\begin{figure*}[t!]
\centering
\includegraphics[width=0.9\textwidth, angle=0]{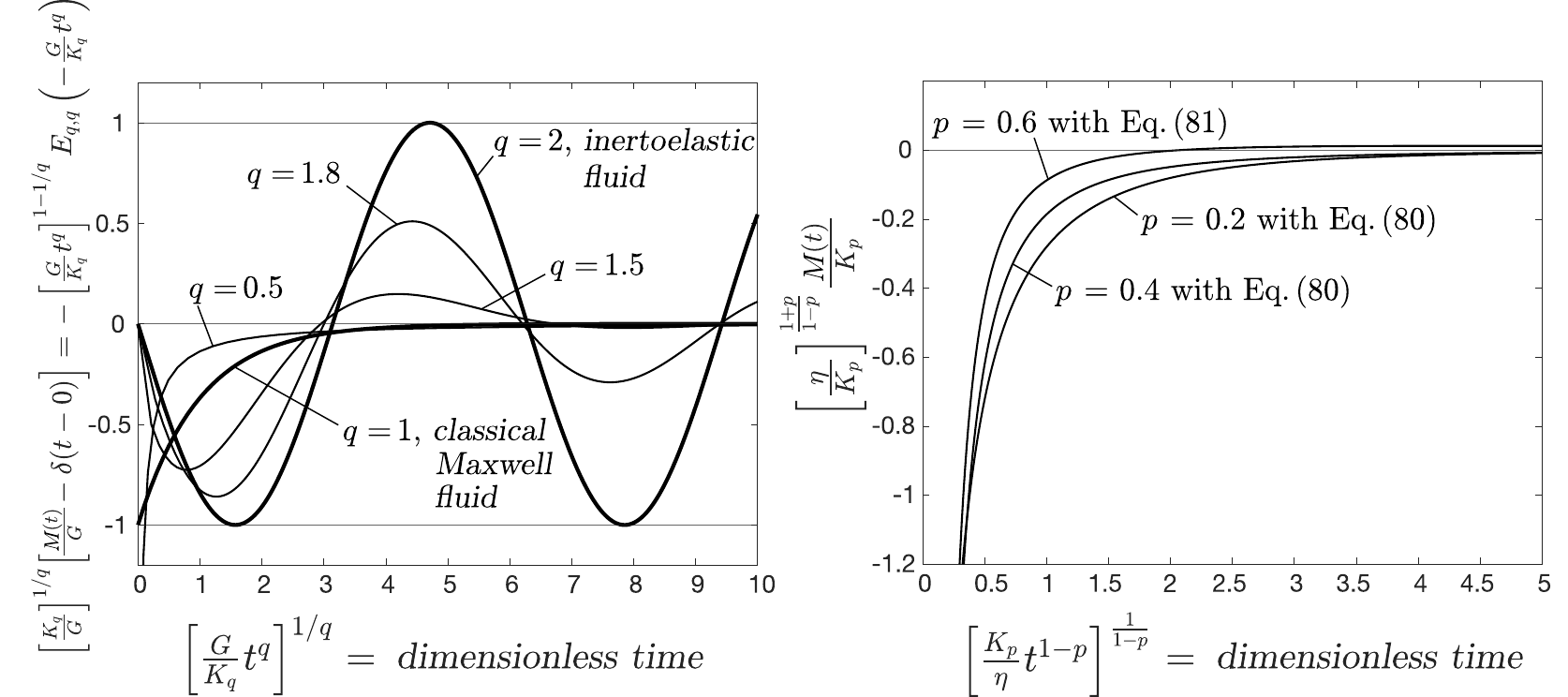}
\caption{Left: Normalized finite part of the memory function, {\large $\left[ \frac{K_q}{G} \right]^{\nicefrac{\text{1}}{q}}\Big[ \frac{M(t)}{G}$} $-\delta(t-\text{0})${\large $\Big]$} of the spring -- Scott-Blair in-series fluid for values $q=$ 0.5, 1, 1.5, 1.8 and 2 as a function of the dimensionless time {\large $\Big[\frac{G}{K_q}$}$t^q${\large $\Big]^{\nicefrac{\text{1}}{q}}$}. For $q\leq$ 1 the time-response functions exhibit a monotonically decreasing behavior; whereas for $q>$ 1, they exhibit an oscillatory behavior capable to capture inertia effects. Right: Normalized memory function of the springpot -- dashpot in-series fluid. For $p=$ 0.2 and $p=$ 0.4, Eq. \eqref{eq:Eq80} was used; whereas for $p=$ 0.6, Eq. \eqref{eq:Eq81} was used.}
\label{fig:Fig04}
\end{figure*}
\begin{align}\label{eq:Eq79}
E_{\text{2, 2}}\left( -\omega_R^{\text{2}} t^{\text{2}}  \right) = &\frac{\sinh(\operatorname{i}\omega_R t)}{\operatorname{i}\omega_R t} = \\ \nonumber
&\frac{\text{1}}{\omega_R t} \frac{e^{\operatorname{i}\omega_R t}-e^{-\operatorname{i}\omega_R t}}{\text{2}\operatorname{i}}= \frac{\text{1}}{\omega_R t} \sin(\omega_R t)\text{,}
\end{align}
Eq. \eqref{eq:Eq78} reduces to $M(t)=G\left[ \delta(t-\text{0}) - \omega_R \sin(\omega_R t) \right] $, which is the memory function of a spring -- inerter connected in-series \citep{Makris2017}. Figure \ref{fig:Fig04} (left) plots the normalized finite part of Eq. \eqref{eq:Eq75}, {\large $\frac{M(t)}{G}$} $-\delta(t-\text{0})$, for various values of $q$ as a function of the dimensionless time {\large $\Big[\frac{G}{K_q}$}$t^q${\large $\Big]^{\nicefrac{\text{1}}{q}}$} and shows that when $q>$ 1, the memory function offered by Eq. \eqref{eq:Eq75} is capable to capture inertia effects in the rheological network as these manifested in high-frequency microrheology \citep{DominguezGarciaCardinauxBertsevaForroScheffoldJeney2014, IndeiSchieberCordoba2012}.

\noindent 2. \textit{Springpot -- Dashpot in-series} $(q=\text{1}$, $r=q-p =\text{1}-p\text{, } K_q=\eta)$. For this case where $q=\text{1}$ and $\text{0} \leq p \leq \text{1}$, we first use Eq. \eqref{eq:Eq73}
\begin{align}\label{eq:Eq80}
M(t)=&K_p \Bigg[ \frac{\mathrm{d}^p}{\mathrm{d}t^p} \delta(t-\text{0}) - \\ \nonumber
&\frac{K_p}{\eta}\frac{\text{1}}{t^{\text{2}p}}  E_{\text{1}-p\text{, }\text{1}-\text{2}p}\left( -\frac{K_p}{\eta}t^{\text{1}-p} \right) \Bigg] \text{,} \enskip \text{0} \leq p \leq \text{1}
\end{align}
Eq. \eqref{eq:Eq80} is valid for values of $\text{0} \leq p <$ {\large $\nicefrac{\text{1}}{\text{2}}$}. In the event that $p \geq$ {\large $\nicefrac{\text{1}}{\text{2}}$}, the second index of the Mittag-Leffler function remains negative and we need to use the next expression for the memory function given by Eq. \eqref{eq:Eq74}
\begin{align}\label{eq:Eq81}
&M(t)=K_p \Bigg[ \frac{\mathrm{d}^p}{\mathrm{d}t^p} \delta(t-\text{0}) -\frac{K_p}{\eta}  \frac{\mathrm{d}^{\text{2}p-\text{1}}}{\mathrm{d}t^{\text{2}p-\text{1}}} \delta(t-\text{0})  - \\ \nonumber
& \left(\frac{K_p}{\eta} \right)^{\text{2}} \frac{\text{1}}{t^{\text{3}p-\text{1}}}  E_{\text{1}-p\text{, }\text{2}-\text{3}p}\left( -\frac{K_p}{\eta}t^{\text{1}-p} \right) \Bigg] \text{,} \enskip \text{0} \leq p \leq \text{1}
\end{align}
Figure \ref{fig:Fig04} (right) plots the normalized memory function of the springpot -- dashpot in-series fluid {\large $\left[ \frac{\eta}{K_p}\right]^{\frac{\text{1}+p}{\text{1}-p}} \frac{M(t)}{K_p}$} for various values of $p$ as a function of the dimensioless time {\large $\Big[\frac{K_p}{\eta}$}$t^{\text{1}-p}${\large $\Big]^{\frac{\text{1}}{\text{1}-p}}$}. For $p=$ 0.2 and $p=$ 0.4, Eq. \eqref{eq:Eq80} was used; whereas for $p=$ 0.6, Eq. \eqref{eq:Eq81} was used. 

The complex dynamic compliance $\mathcal{J}(s)=$ {\large $\nicefrac{\text{1}}{\mathcal{G}(s)}$} derives directly from Eq. \eqref{eq:Eq67} in which $r=q-p$ and $\lambda_r=$ {\large $\nicefrac{K_q}{K_p}$}
\begin{equation}\label{eq:Eq82}
\mathcal{J}(s)=\frac{\text{1}}{K_q} \frac{\text{1}+\lambda_r s^{q-p}}{s^q}=\frac{\text{1}}{K_p}\frac{\text{1}}{s^p} + \frac{\text{1}}{K_q}\frac{\text{1}}{s^q}
\end{equation}
The impulse fluidity (impulse response function), $\phi(t)$, is the superposition of the impulse fluidities of the two Scott-Blair elements connected in-series
\begin{align}\label{eq:Eq83}
\phi(t)= & \mathcal{L}^{-\text{1}} \left\lbrace \mathcal{J}(s) \right\rbrace = \Bigg[ \frac{\text{1}}{K_p} \frac{\text{1}}{\Gamma(p)} \frac{\text{1}}{t^{\text{1}-p}} + \\ \nonumber
&   \frac{\text{1}}{K_q} \frac{\text{1}}{\Gamma(q)} \frac{\text{1}}{t^{\text{1}-q}} \Bigg]U(t-\text{0}) \text{,} \enskip \text{0}<p<q \in \mathbb{R}^+
\end{align}

\noindent \textbf{Special Cases:} 1. \textit{Spring -- Scott-Blair element in-series} $(p=\text{0}$, $r=q-p=q \in \mathbb{R}^+ \text{, } K_p=G)$. In the case where $p=\text{0}$, the first term in the bracket of Eq. \eqref{eq:Eq83} becomes the Dirac delta function $\delta(t-\text{0})$ according to Eq. \eqref{eq:Eq31} \citep{GelfandShilov1964}. Consequently, for $p=\text{0}$ and $K_p=G$ in association with Eq. \eqref{eq:Eq31}, Eq. \eqref{eq:Eq83} yields
\begin{equation}\label{eq:Eq84}
\phi(t)=\frac{\text{1}}{G}\delta(t-\text{0})+\frac{\text{1}}{K_q}\frac{\text{1}}{\Gamma(q)}\frac{\text{1}}{t^{\text{1}-q}}U(t-\text{0}) \text{,} \enskip q \in \mathbb{R}^+
\end{equation}

\noindent which is the superposition of the impulse fluidities of the Hookean spring and that of the Scott-Blair (springpot) element shown in Table \ref{tab:Table2}. When $q=\text{2}$, $K_q=K_\text{2}=m_R$, Eq. \eqref{eq:Eq84} gives $\phi(t)=$ {\large $\frac{\text{1}}{G}$}$\delta(t-\text{0})+${\large $\frac{\text{1}}{m_R}$}$tU(t-\text{0})$, which is the impulse fluidity of a spring -- inerter in-series connection \citep{Makris2017}.

\noindent 2. \textit{Springpot -- Dashpot in-series} $(q=\text{1}$, $r=q-p =\text{1}-p\text{, } K_q=\eta)$. In this case where $q=\text{1}$, Eq. \eqref{eq:Eq83} yields 
\begin{equation}\label{eq:Eq85}
\phi(t)=\left[ \frac{\text{1}}{K_p}\frac{\text{1}}{\Gamma(p)}\frac{\text{1}}{t^{\text{1}-p}} +\frac{\text{1}}{\eta} \right]U(t-\text{0})
\end{equation}
which is the superposition of the impulse fluidities of the springpot element and that of a dashpot. For the classical limit when $p=\text{0}$, $K_p=G$, $q=\text{1}$ and $K_q=\eta$, Eq. \eqref{eq:Eq84} yields $\phi(t)=$ {\large $\frac{\text{1}}{G}$}$\delta(t-\text{0})+$ {\large $\frac{\text{1}}{\eta}$}$U(t-\text{0})$ which is the impulse fluidity (impulse response function) of the classical Maxwell model shown in Table \ref{tab:Table2}.

The complex dynamic viscosity (impedance), $\eta(s)$, of the generalized Maxwell fluid derives directly from Eq. \eqref{eq:Eq67}, since $\eta(s)=$ {\large $\nicefrac{\mathcal{G}(s)}{s}$}
\begin{equation}\label{eq:Eq86}
\eta(s)=K_p \frac{s^{q-\text{1}}}{s^r + \nicefrac{K_p}{K_q}} = K_p \frac{s^{r-(\text{1}-p)}}{s^r + \nicefrac{K_p}{K_q}} \text{,} \enskip r=q-p>\text{0}
\end{equation}
The relaxation modulus, $G(t)$, is the inverse Laplace transform of Eq. \eqref{eq:Eq86} 
\begin{align}\label{eq:Eq87}
G(t)=& \mathcal{L}^{-\text{1}}\left\lbrace \mathcal{\eta}(s) \right\rbrace=\mathcal{L}^{-\text{1}}\left\lbrace K_p \frac{s^{r-(\text{1}-p)}}{s^r + \nicefrac{K_p}{K_q}} \right\rbrace = \\ \nonumber
& K_p \, \frac{\text{1}}{t^p} \, E_{r\text{, }\text{1}-p}\left( -\frac{K_p}{K_q}t^r \right) \text{,} \enskip t>\text{0}
\end{align}
and by replacing $r=q-p$,
\begin{equation}\label{eq:Eq88}
G(t)=K_p \, \frac{\text{1}}{t^p} \, E_{q-p\text{, }\text{1}-p} \left( -\frac{K_p}{K_q}t^{q-p} \right) \text{,} \enskip \text{0}<p<q \in \mathbb{R}^+
\end{equation}
The expression given by Eq. \eqref{eq:Eq88} has been presented by \citet{Friedrich1991, SchiesselMetzlerBlumenNonnenmacher1995, PaladeVerneyAttane1996} and was employed by \citet{JaishankarMcKinley2013}

\begin{figure*}[t!]
\centering
\includegraphics[width=0.9\textwidth, angle=0]{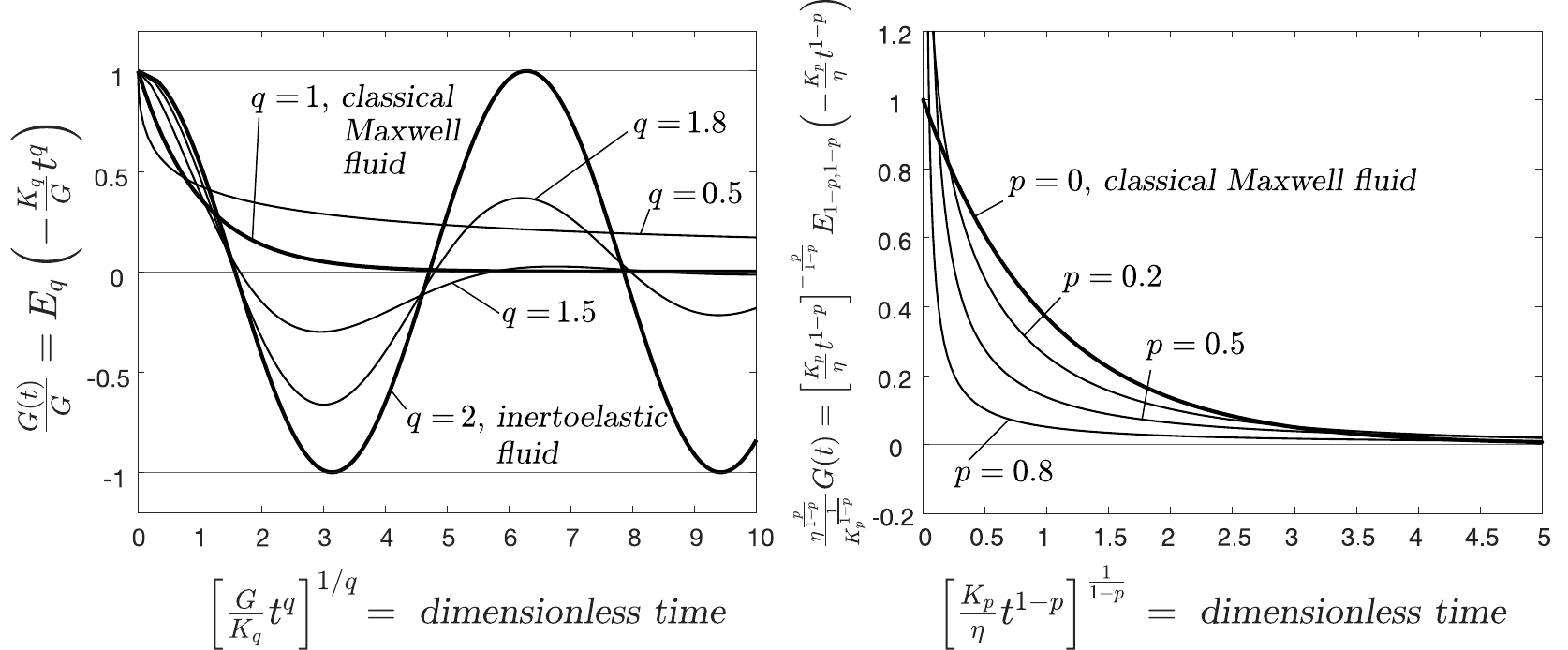}
\caption{Left: Normalized relaxation modulus {\large $\frac{G(t)}{G}$} of the spring -- Scott-Blair in-series fluid for values $q=$ 0.5, 1, 1.5, 1.8 and 2 as a function of the dimensionless time {\large $\Big[\frac{G}{K_q}$}$t^q${\large $\Big]^{\nicefrac{\text{1}}{q}}$}. For $q\leq$ 1, the time-response functions exhibit a monotonically decreasing behavior; whereas for $q>$ 1, they exhibit an oscillatory behavior capable to capture inertia effects. Figure \ref{fig:Fig04} (left) has been presented by \citet{GlockleNonnenmacher1994} by merely allowing $q$ to assume values larger than 1, about a decade before the concept of the inerter and its equivalence to the electric capacitor was established by \citet{Smith2002}. Right: Normalized relaxation modulus of the springpot -- dashpot in-series fluid for values of $p=$ 0, 0.2, 0.5 and 0.8 as a function of the dimensionless time {\large $\Big[\frac{K_p}{\eta}$}$t^{\text{1}-p}${\large $\Big]^{\frac{\text{1}}{\text{1}-p}}$} with 0 $\leq p \leq$ 1.}
\label{fig:Fig05}
\end{figure*}

\noindent \textbf{Special Cases:} 1. \textit{Spring -- Scott-Blair element in-series} $(p=\text{0}$, $r=q-p=q \in \mathbb{R}^+ \text{, } K_p=G)$. In this case where $p=\text{0}$, Eq. \eqref{eq:Eq88} reduces to
\begin{align}\label{eq:Eq89}
G(t)=&G \, E_{q\text{, 1}} \left( -\frac{G}{K_q}t^q \right) = \\ \nonumber
&G \, E_q \left( -\frac{G}{K_q}t^q \right)  \text{,} \enskip q \in \mathbb{R}^+
\end{align}
The result of Eq. \eqref{eq:Eq89} has been presented by \citet{Koeller1984, SchiesselMetzlerBlumenNonnenmacher1995, MainardiSpada2011}. When $q=\text{1}$, the Mittag-Leffler function $E_{\text{1}}\Big(-$ {\large $\frac{G}{\eta}$}$t\Big)$ reduces to the exponential function $e^-${\large $^{\nicefrac{t}{\lambda}}$} where $\lambda=$ {\large $\nicefrac{\eta}{G}$} $=$ relaxation time; and Eq. \eqref{eq:Eq89} gives the relaxation modulus of the classical Maxwell model $G(t)=Ge^-${\large $^{\nicefrac{t}{\lambda}}$}. When $q=\text{2}$, $K_q=K_{\text{2}}=m_R$ and {\large $\frac{G}{K_{\text{2}}}$} $=$ {\large $\frac{G}{m_R}$} $=\omega_R^{\text{2}}$. By virtue of the identity $E_{\text{2, 1}}\left( -\omega_R^{\text{2}}t^{\text{2}} \right)=E_{\text{2}}\left( -\omega_R^{\text{2}}t^{\text{2}} \right)=\cos(\omega_R^{\text{2}}t)$, Eq. \eqref{eq:Eq89} reduces to $G(t)=G\cos(\omega_R t)$ which is the relaxation modulus of a spring -- inerter connected in-series \citep{Makris2017}. The emerging of the $\cos$ine function (oscillatory behavior) when $q=$ 2 has been reported by \citet{GlockleNonnenmacher1994} after examining the solutions of the Fox H-function that is related to the Mittag-Leffler function. The oscillatory behavior of the relaxation modulus, $G(t)$, when $q=$ 2 is the result of the continuous exchange of potential and kinetic energies between the spring $(p=\text{0})$ and the inerter $(q=\text{2})$. Figure \ref{fig:Fig05} (left) plots the relaxation modulus given by Eq. \eqref{eq:Eq89} for various values of $q$ as a function of the dimensionless time {\large $\Big[\frac{G}{K_q}$}$t^q${\large $\Big]^{\nicefrac{\text{1}}{q}}$}.

\noindent 2. \textit{Springpot -- Dashpot in-series} $(q=\text{1}$, $r=q-p =\text{1}-p\text{, } K_q=\eta)$. In this case where $q=\text{1}$, Eq. \eqref{eq:Eq88} reduces to
\begin{equation}\label{eq:Eq90}
G(t)=K_p \, \frac{\text{1}}{t^p} \, E_{\text{1}-p\text{, }\text{1}-p}\left( -\frac{K_p}{\eta} t^{\text{1}-p} \right) \text{,} \enskip \text{0} \leq p \leq \text{1} 
\end{equation}
Figure \ref{fig:Fig05} (right) plots the normalized relaxation modulus, $G(t)$, of the springpot -- dashpot in-series element for various values of $p$ $(\text{0} \leq p \leq \text{1}) $.

The complex dynamic fluidity (admittance), $\phi(s)$, of the generalized fractional Maxwell fluid is the inverse of the complex dynamic viscosity given by Eq. \eqref{eq:Eq86}
\begin{align}\label{eq:Eq91}
\phi(s)=&\frac{\text{1}}{K_p} \frac{s^r+\nicefrac{K_p}{K_q}}{s^{q-\text{1}}}= \\ \nonumber
& \frac{\text{1}}{K_p} s^{\text{1}-p} + \frac{\text{1}}{K_q} s^{\text{1}-q} \text{,} \enskip r=q-p>\text{0}
\end{align}
The impulse strain-rate response function of the generalized fractional Maxwell fluid, $\psi(t)$, is the superposition of the impulse strain-rate response functions of the two Scott-Blair elements connected in-series
\begin{align}\label{eq:Eq92}
\psi(t)=&\mathcal{L}^{-\text{1}} \left\lbrace \phi(s) \right\rbrace = \Bigg[ \frac{\text{1}}{K_p} \frac{\text{1}}{\Gamma(-\text{1}+p)} \frac{\text{1}}{t^{\text{2}-p}} + \\ \nonumber
& \frac{\text{1}}{K_q} \frac{\text{1}}{\Gamma(-\text{1}+q)} \frac{\text{1}}{t^{\text{2}-q}} \Bigg] U(t-\text{0}) \text{,} \enskip \text{0}<p<q \in \mathbb{R}^+
\end{align}

\noindent \textbf{Special Cases:} 1. \textit{Spring -- Scott-Blair element in-series} $(p=\text{0}$, $r=q-p=q \in \mathbb{R}^+ \text{, } K_p=G)$. In this case where $p=\text{0}$, the first term in the bracket of Eq. \eqref{eq:Eq92} becomes equal to the first derivative of the Dirac delta function, {\large $\frac{\mathrm{d}\delta(t-\text{0})}{\mathrm{d}t}$}, according to Eq. \eqref{eq:Eq31}. Consequently for $p=\text{0}$ and $K_p=G$, in association with Eq. \eqref{eq:Eq31}, Eq. \eqref{eq:Eq92} yields
\begin{equation}\label{eq:Eq93}
\psi(t)=\frac{\text{1}}{G} \frac{\mathrm{d}\delta(t-\text{0})}{\mathrm{d}t} + \frac{\text{1}}{K_q} \frac{\text{1}}{\Gamma(-\text{1}+q)} \frac{\text{1}}{t^{\text{2}-q}} U(t-\text{0})
\end{equation}

\noindent 2. \textit{Springpot -- Dashpot in-series:} $(q=\text{1}$, $r=q-p =\text{1}-p\text{, } K_q=\eta)$. In this case where $q=\text{1}$, the second term in the bracket of Eq. \eqref{eq:Eq92} becomes equal to the Dirac delta function, $\delta(t-\text{0})$, according to Eq. \eqref{eq:Eq31}. Consequently for $q=\text{1}$ and $K_q=\eta$, in association with Eq. \eqref{eq:Eq31}, Eq. \eqref{eq:Eq92} yields
\begin{equation}\label{eq:Eq94}
\psi(t)=\frac{\text{1}}{K_p} \frac{\text{1}}{\Gamma(-\text{1}+p)} \frac{\text{1}}{t^{\text{2}-p}}U(t-\text{0}) + \frac{\text{1}}{\eta}\delta(t-\text{0})
\end{equation} 
 
The complex creep function, $\mathcal{C}(\omega)$, of the generalized fractional Maxwell fluid derives directly from Eq. \eqref{eq:Eq82} given that $\mathcal{C}(s)=$ {\large $\nicefrac{\mathcal{J}(s)}{s}$}
\begin{equation}\label{eq:Eq95}
\mathcal{C}(s)=\frac{\text{1}}{K_q} \frac{\text{1}+\lambda_r s^{q-p}}{s^{\text{1}+q}} = \frac{\text{1}}{K_p} \frac{\text{1}}{s^{\text{1}+p}}+ \frac{\text{1}}{K_q} \frac{\text{1}}{s^{\text{1}+q}}
\end{equation}

\begin{sidewaystable*}
\centering
\caption{Frequency-response functions and the corresponding causal time-response functions of the generalized fractional derivative Maxwell fluid and of its special cases.}
\setlength{\tabcolsep}{2pt}
{\renewcommand{\arraystretch}{1.5}
\begin{tabularx}{\textheight}{>{\centering}p{0.13\textheight}>{\centering}p{0.29\textheight}>{\centering}p{0.29\textheight} >{\centering}p{0.29\textheight}}
\hline \hline
	&  \thead{\textbf{Generalized Fractional Derivative} \\ \textbf{Maxwell Fluid}} & \thead{\textbf{Spring -- Scott-Blair} \\ \textbf{In-Series Connection}} & \thead{\textbf{Springpot -- Dashpot} \\ \textbf{In-Series Connection}}
	\tabularnewline
	& \includegraphics[scale=0.25]{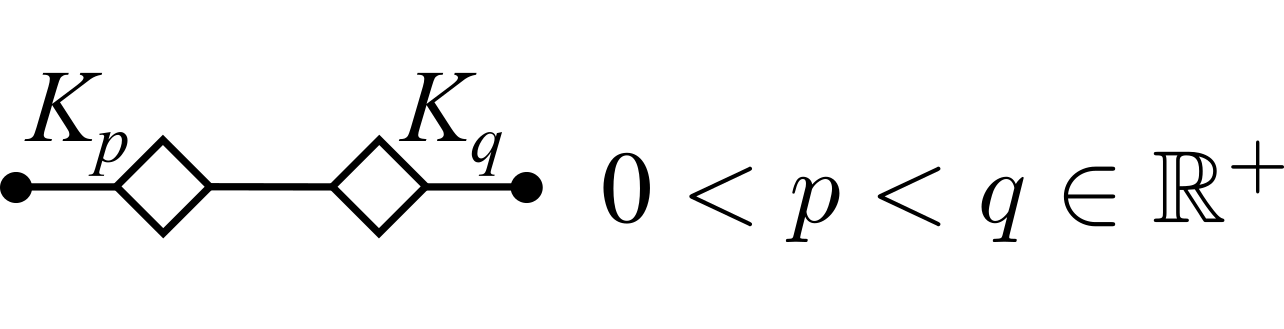} & \includegraphics[scale=0.25]{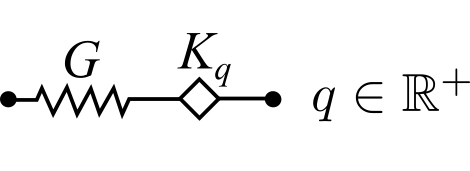} & \includegraphics[scale=0.25]{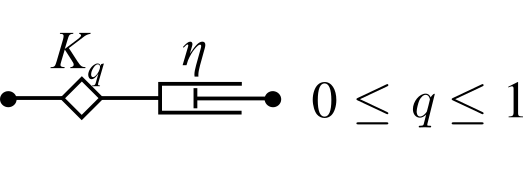}
	\tabularnewline 
	\thead{Constitutive \\ Equation} & $\tau(t)+\cfrac{K_q}{K_p}\cfrac{\mathrm{d}^{q-p}}{\mathrm{d}t^{q-p}}\tau(t)=K_q\cfrac{\mathrm{d}^q\gamma(t)}{\mathrm{d}t^q}$ & $\tau(t)+\cfrac{K_q}{G}\cfrac{\mathrm{d}^q\tau(t)}{\mathrm{d}t^q} = K_q \cfrac{\mathrm{d}^q\gamma(t)}{\mathrm{d}t^q}$ &  $\tau(t)+\cfrac{\eta}{K_q}\cfrac{\mathrm{d}^{\text{1}-q}}{\mathrm{d}t^{\text{1}-q}}\tau(t)=\eta\cfrac{\mathrm{d}\gamma(t)}{\mathrm{d}t}$
	\tabularnewline  \hline 
	\thead{Complex \\ Dynamic Modulus \\ $\mathcal{G}(\omega)=\frac{\tau(\omega)}{\gamma(\omega)}$}  & $K_q\cfrac{(\operatorname{i}\omega)^q}{\text{1}+\frac{K_q}{K_p}(\operatorname{i}\omega)^{q-p}}$ & $K_q\cfrac{(\operatorname{i}\omega)^q}{\text{1}+\frac{K_q}{G}(\operatorname{i}\omega)^q}$  &  $\eta\cfrac{\operatorname{i}\omega}{\text{1}+\frac{\eta}{K_q}(\operatorname{i}\omega)^{\text{1}-q}}$
	\tabularnewline  
	\thead{Complex\\ Dynamic Viscosity \\ $\mathcal{\eta}(\omega)=\frac{\tau(\omega)}{\dt{\gamma}(\omega)}$}  & $K_p \cfrac{(\operatorname{i}\omega)^{q-\text{1}}}{(\operatorname{i}\omega)^{q-p}+\frac{K_p}{K_q}}$ & $G \cfrac{(\operatorname{i}\omega)^q}{\operatorname{i}\omega\left[ (\operatorname{i}\omega)^q +\frac{G}{K_q} \right]}$  & $\cfrac{\eta }{\text{1}+\frac{\eta}{K_q}(\operatorname{i}\omega)^{\text{1}-q}}$
	\tabularnewline  
	\thead{Complex \\ Dynamic Compliance \\ $\mathcal{J}(\omega)=\frac{\text{1}}{\mathcal{G}(\omega)}=\frac{\gamma(\omega)}{\tau(\omega)}$} & $\cfrac{\text{1}}{K_p}\cfrac{\text{1}}{(\operatorname{i}\omega)^p}+\cfrac{\text{1}}{K_q}\cfrac{\text{1}}{(\operatorname{i}\omega)^q}$ & $\cfrac{\text{1}}{G}+\cfrac{\text{1}}{K_q}\cfrac{\text{1}}{(\operatorname{i}\omega)^q}$ & $\cfrac{\text{1}}{\eta} \left[ \pi \delta(\omega-\text{0}) - \operatorname{i}\cfrac{\text{1}}{\omega} \right] + \cfrac{\text{1}}{K_q} \cfrac{\text{1}}{(\operatorname{i}\omega)^q}$
	\tabularnewline
	\thead{Complex  \\ Creep Function \\ $\mathcal{C}(\omega)=\frac{\gamma(\omega)}{\dt{\tau}(\omega)}$}  & $\cfrac{\text{1}}{K_p}\cfrac{\text{1}}{(\operatorname{i}\omega)^{\text{1}+p}}+\cfrac{\text{1}}{K_q}\cfrac{\text{1}}{(\operatorname{i}\omega)^{\text{1}+q}}$  & $\cfrac{\text{1}}{G}\left[ \pi \delta(\omega-\text{0}) -\operatorname{i}\cfrac{\text{1}}{\omega} \right]+\cfrac{\text{1}}{K_q}\cfrac{\text{1}}{(\operatorname{i}\omega)^{q+\text{1}}}$  & $\cfrac{\text{1}}{\eta} \left[ -\cfrac{\text{1}}{\omega^{\text{2}}} + \operatorname{i}\pi \cfrac{\mathrm{d}\delta(\omega-\text{0})}{\mathrm{d}\omega} \right] +\cfrac{\text{1}}{K_q} \cfrac{\text{1}}{(\operatorname{i}\omega)^{q+\text{1}}}$
	\tabularnewline 
	\thead{ Complex \\ Dynamic Fluidity\\ $\mathcal{\phi}(\omega)=\frac{\text{1}}{\eta(\omega)}=\frac{\dt{\gamma}(\omega)}{\tau(\omega)}$} & $\cfrac{\text{1}}{K_p}(\operatorname{i}\omega)^{\text{1}-p}+\cfrac{\text{1}}{K_q}(\operatorname{i}\omega)^{\text{1}-q}$ & $\cfrac{\text{1}}{G}\operatorname{i}\omega+\cfrac{\text{1}}{K_q}(\operatorname{i}\omega)^{\text{1}-q}$ & $\cfrac{\text{1}}{K_q}(\operatorname{i}\omega)^{\text{1}-q}+\cfrac{\text{1}}{\eta}$
	\tabularnewline  \hline 
	\thead{Memory Function \\ $M(t)$} & \thead{$ \cfrac{K_p}{t^{p+\text{1}}} \Bigg[ \cfrac{\text{1}}{\Gamma(-p)} U(t-\text{0})- \quad \quad \quad \quad \quad \quad$ \\ $ \quad \quad \quad \quad \cfrac{K_p}{K_q}\cfrac{\text{1}}{t^{p-q}} E_{q-p\text{, }q-\text{2}p}\left( -\cfrac{K_p}{K_q}t^{q-p} \right) \Bigg]$} &  $G\left[ \delta(t-\text{0})-\cfrac{G}{K_q}\cfrac{\text{1}}{t^{\text{1}-q}}E_{q\text{, }q}\left( -\cfrac{G}{K_q}t^q \right) \right]$ & \thead{$\cfrac{K_q}{t^{q+\text{1}}}  \Bigg[ \cfrac{\text{1}}{\Gamma(-q)} U(t-\text{0})- \quad \quad \quad \quad \quad \quad \quad \quad \quad $ \\ $\quad \quad \cfrac{K_q}{\eta}\cfrac{\text{1}}{t^{q-\text{1}}}  E_{\text{1}-q\text{, }\text{1}-\text{2}q}\left( -\cfrac{K_q}{\eta}t^{\text{1}-q} \right) \Bigg] $}
	\tabularnewline 
	\thead{Relaxation Modulus \\ $G(t)$} & $K_p \, \cfrac{\text{1}}{t^p} \, E_{q-p\text{, }\text{1}-p}\left( -\cfrac{K_p}{K_q}t^{q-p}\right)  $ &  $G \, E_q\left( -\cfrac{G}{K_q}t^q \right)$ & $K_q \, \cfrac{\text{1}}{t^q} \, E_{\text{1}-q\text{, }\text{1}-q}\left( -\cfrac{K_q}{\eta}t^{\text{1}-q}\right)  $
	\tabularnewline  
	\thead{Impulse Fluidity \\ $\phi(t)$} & $\left[ \cfrac{\text{1}}{K_p} \cfrac{\text{1}}{\Gamma(p)} \cfrac{\text{1}}{t^{\text{1}-p}} + \cfrac{\text{1}}{K_q} \cfrac{\text{1}}{\Gamma(q)} \cfrac{\text{1}}{t^{\text{1}-q}} \right] U(t-\text{0})$ &  $\cfrac{\text{1}}{G} \, \delta(t-\text{0})+\cfrac{\text{1}}{K_q}\cfrac{\text{1}}{\Gamma(q)}\cfrac{\text{1}}{t^{\text{1}-q}}U(t-\text{0})$ & $\left[ \cfrac{\text{1}}{K_q} \cfrac{\text{1}}{\Gamma(q)} \cfrac{\text{1}}{t^{\text{1}-q}} + \cfrac{\text{1}}{\eta} \right]U(t-\text{0})$
	\tabularnewline  
	\thead{Creep Compliance \\ $J(t)$}	& $\left[ \cfrac{\text{1}}{K_p} \cfrac{t^p}{\Gamma(p+\text{1})} + \cfrac{\text{1}}{K_q} \cfrac{t^q}{\Gamma(q+\text{1})}  \right] U(t-\text{0})$  &   $\left[\cfrac{\text{1}}{G}+\cfrac{\text{1}}{K_q}\cfrac{t^q}{\Gamma(q+\text{1})}\right]U(t-\text{0})$ &  $\left[ \cfrac{\text{1}}{K_q} \cfrac{\text{1}}{\Gamma(q+\text{1})}t^q +\cfrac{\text{1}}{\eta}t  \right]U(t-\text{0})$
	\tabularnewline 
	\thead{Impulse Strain-rate \\ Response Function \\ $\psi(t)$}	& \scriptsize{$\left[ \cfrac{\text{1}}{K_p} \cfrac{\text{1}}{\Gamma(-\text{1}+p)} \cfrac{\text{1}}{t^{\text{2}-p}} + \cfrac{\text{1}}{K_q} \cfrac{\text{1}}{\Gamma(-\text{1}+q)} \cfrac{\text{1}}{t^{\text{2}-q}} \right] U(t-\text{0})$} &   $\cfrac{\text{1}}{G}\cfrac{\mathrm{d}\delta(t-\text{0})}{\mathrm{d}t}+\cfrac{\text{1}}{K_q}\cfrac{\text{1}}{\Gamma(-\text{1}+q)}\cfrac{\text{1}}{t^{\text{2}-q}}U(t-\text{0})$ & $ \cfrac{\text{1}}{K_q} \cfrac{\text{1}}{\Gamma(-\text{1}+q)} \cfrac{\text{1}}{t^{\text{2}-q}}U(t-\text{0}) + \cfrac{\text{1}}{\eta}\delta(t-\text{0}) $
	\tabularnewline  \hline \hline
\end{tabularx}}
\label{tab:Table3}
\end{sidewaystable*}

\noindent The creep compliance (retardation function), $J(t)$, is the superposition of the creep compliances of the two Scott-Blair elements connected in-series.
\begin{align}\label{eq:Eq96}
J(t)=&\mathcal{L}^{-\text{1}} \left\lbrace \mathcal{C}(s) \right\rbrace =   \\ \nonumber
& \left[ \frac{\text{1}}{K_p}\frac{t^p}{\Gamma(p+\text{1})} +\frac{\text{1}}{K_q}\frac{t^q}{\Gamma(q+\text{1})} \right]U(t-\text{0})
\end{align}
The expression given by Eq. \eqref{eq:Eq96} has been presented by \citet{Friedrich1991, SchiesselMetzlerBlumenNonnenmacher1995, JaishankarMcKinley2013, Hristov2019}.

\noindent \textbf{Special Cases:} 1. \textit{Spring -- Scott-Blair element in-series} $(p=\text{0}$, $r=q-p=q \in \mathbb{R}^+ \text{, } K_p=G)$. In this case where $p=\text{0}$, Eq. \eqref{eq:Eq96} reduces to
\begin{equation}\label{eq:Eq97}
J(t)=\left[ \frac{\text{1}}{G} + \frac{\text{1}}{K_q}\frac{t^q}{\Gamma(q+\text{1})} \right]U(t-\text{0})
\end{equation}
The result of Eq. \eqref{eq:Eq97} has been presented by \citet{Koeller1984, SchiesselMetzlerBlumenNonnenmacher1995, MainardiSpada2011}.

\noindent 2. \textit{Springpot -- Dashpot in-series} $(q=\text{1}$, $r=q-p =\text{1}-p\text{, } K_q=\eta)$. In this case where $p=\text{1}$, Eq. \eqref{eq:Eq96} reduces to
\begin{equation}\label{eq:Eq98}
J(t)=\left[ \frac{\text{1}}{K_p}\frac{t^p}{\Gamma(p+\text{1})} + \frac{\text{1}}{\eta}t \right]U(t-\text{0})
\end{equation}
For the classical limit when $p=\text{0}$, $K_p=G$, $q=\text{1}$ and $K_q=\eta$, Eq. \eqref{eq:Eq96} yields $J(t)=$ {\large $\Big[ \frac{\text{1}}{G} +\frac{\text{1}}{\eta}$}$t${\large $\Big]$}$U(t-\text{0})$ which is the creep compliance of the classical Maxwell fluid shown in Table \ref{tab:Table2}.

The five causal time-response functions of the generalized fractional Maxwell fluid together with the time-response functions for the special cases of the spring -- Scott-Blair element in-series connection $(p=\text{0})$ and the springpot -- dashpot in-series connection $(q=\text{1})$ are summarized in Table \ref{tab:Table3}.

\section{Time-response function of the generalized fractional Kelvin-Voigt element}
The parallel connection of the two Scott-Blair elements shown in Figure \ref{fig:Fig03} (right) exhibits a solid-like behavior only when $p=\text{0}$ or $q=\text{0}$. In any other situation where $p>\text{0}$ and $q>\text{0}$, the generalized fractional Kelvin-Voigt element shown in Figure \ref{fig:Fig03} (right) exhibits a fluid-like behavior since it results in infinite deformation under a static load.
In view of this behavior the term ``generalized'' fractional Kelvin-Voigt element is used for the viscoelastic model shown in Figure \ref{fig:Fig03} (right) with constitutive law
\begin{equation}\label{eq:Eq99}
\tau(t)=K_p\frac{\mathrm{d}^p\gamma(t)}{\mathrm{d}t^p}+K_q\frac{\mathrm{d}^q\gamma(t)}{\mathrm{d}t^q} \text{,} \enskip p\text{, } q \in \mathbb{R}^+
\end{equation}
When $p=$ 0, the generalized fractional Kelvin-Voigt element given by Eq. \eqref{eq:Eq99} reduces to a spring -- Scott-Blair element parallel connection --- a model that was proposed by \citet{SukiBarabasiLutchen1994} to express the pressure--volume relation of the lung tissue viscoelastic behavior of human and selective animal lungs. The same spring -- Scott-Blair element was subsequently used by \citet{PuigdeMoralesMarinkovicTurnerButlerFredbergSuresh2007} to model the viscoelastic behavior of the human red blood cells. When $(\text{0} \leq p \leq \text{1 and } q =\text{1})$  the generalized fractional Kelvin-Voigt element given by Eq. \eqref{eq:Eq99} reduces to a springpot -- dashpot parallel connection --- a model that has been used to capture the high-frequency behavior of semiflexible polymer networks \citep{GittesMacKintosh1998, AtakhorramiMizunoKoenderinkLiverpoolMacKintoshSchmidt2008, DominguezGarciaCardinauxBertsevaForroScheffoldJeney2014}.

Because of the parallel arrangement of the two Scott-Blair elements, the memory function, $M(t)$, of the generalized fractional Kelvin-Voigt element is the summation of the memory functions of the two individual Scott-Blair elements given by Eq. \eqref{eq:Eq36}
\begin{align}\label{eq:Eq100}
M(t)=&\left[ \frac{K_p}{\Gamma(-p)}\frac{\text{1}}{t^{p+1}} + \frac{K_q}{\Gamma(-q)}\frac{\text{1}}{t^{q+1}} \right]U(t-\text{0}) = \\ \nonumber
&K_p\frac{\mathrm{d}^p\delta(t-\text{0})}{\mathrm{d}t^p}+K_q\frac{\mathrm{d}^q\delta(t-\text{0})}{\mathrm{d}t^q}
\end{align}

\noindent \textbf{Special Cases:} 1. \textit{Spring -- Scott-Blair element in parallel} $(p=\text{0}$, $q \in \mathbb{R}^+ \text{, } K_p=G)$.
\begin{equation}\label{eq:Eq101}
M(t)=G\delta(t-\text{0})+\frac{K_q}{\Gamma(-q)}\frac{\text{1}}{t^{q+\text{1}}}U(t-\text{0})
\end{equation}
\noindent 2. \textit{Springpot -- Dashpot in parallel} $(p \in \mathbb{R}^+$, $q=\text{1}$, $K_q=\eta)$.
\begin{equation}\label{eq:Eq102}
M(t)=\frac{K_p}{\Gamma(-p)}\frac{\text{1}}{t^{p+\text{1}}}U(t-\text{0})+\eta\frac{\mathrm{d}\delta(t-\text{0})}{\mathrm{d}t}
\end{equation}
For the classical limit when $p=\text{0}$, $K_p=G$, $q=\text{1}$ and $K_q=\eta$, Eq. \eqref{eq:Eq100} yields $M(t)=G\delta(t-\text{0})+\eta${\large $\frac{\mathrm{d}\delta(t-\text{0})}{\mathrm{d}t}$} which is the memory function of the classical Kelvin-Voigt solid shown in Table \ref{tab:Table2}. When $q=$ 2 and $K_q=K_{\text{2}}=m_R$, Eq. \eqref{eq:Eq101} yields $M(t)=G\delta(t-\text{0})+m_R${\large $\frac{\mathrm{d}^{\text{2}}\delta(t-\text{0})}{\mathrm{d}t^{\text{2}}}$}, which is the memory function of the inerto-elastic solid \citep{Makris2017}.

The complex dynamic compliance, $\mathcal{J}(\omega)=$ {\large $\nicefrac{\gamma(\omega)}{\tau(\omega)}$}, derives directly from the Fourier transform of Eq. \eqref{eq:Eq99}, $\tau(\omega)=\left[ K_p(\operatorname{i}\omega)^p + K_q(\operatorname{i}\omega)^q\right]\gamma(\omega)$ and by using the Laplace variable $s=\operatorname{i}\omega$
\begin{align}\label{eq:Eq103}
\mathcal{J}(s)=&\frac{\gamma(s)}{\tau(s)}=\frac{\text{1}}{K_p s^p+K_q s^q}= \\ \nonumber
&\frac{\text{1}}{K_p}\frac{\text{1}}{s^p\left( \text{1}+\frac{K_q}{K_p}s^{q-p} \right)} =\frac{\text{1}}{K_q}\frac{\text{1}}{s^p\left( s^r+\frac{K_p}{K_q}\right)} \text{,} \\ \nonumber
&\quad \quad \quad \quad \quad \quad \quad \quad \quad \quad \quad \quad \text{with} \enskip p\text{, }r \in \mathbb{R}^+
\end{align}
where without loss of generality we set $r=q-p>\text{0}$. The inverse Laplace transform of Eq. \eqref{eq:Eq103} is evaluated with the convolution integral given by Eq. \eqref{eq:Eq68}, where
\begin{align}\label{eq:Eq104}
f(t)=&\mathcal{L}^{-\text{1}} \left\lbrace \mathcal{F}(s) \right\rbrace = \mathcal{L}^{-\text{1}} \left\lbrace \frac{\text{1}}{K_q}\frac{\text{1}}{s^p} \right\rbrace = \\ \nonumber
&\frac{\text{1}}{K_q}\frac{\text{1}}{\Gamma(p)}t^{p-\text{1}} \text{,} \enskip p \in \mathbb{R}^+
\end{align}
and $h(t)=\mathcal{L}^{-\text{1}} \left\lbrace \mathcal{H}(s) \right\rbrace $ is given by Eq. \eqref{eq:Eq70}. Substitution of the results of Eqs. \eqref{eq:Eq104} and \eqref{eq:Eq70} into the convolution integral given by Eq. \eqref{eq:Eq68}, the impulse fluidity (impulse response function) of the generalized fractional Kelvin-Voigt element is 
\begin{align}\label{eq:Eq105}
\phi(t)=&\mathcal{L}^{-\text{1}} \left\lbrace \mathcal{J}(s) \right\rbrace = \\ \nonumber
&\frac{\text{1}}{K_q}\frac{\text{1}}{\Gamma(p)}\int_{\text{0}}^{t} (t-\xi)^{p-\text{1}}\xi^{r-\text{1}}E_{r\text{, }r}\left( -\frac{K_p}{K_q}\xi^r \right) \mathrm{d}\xi
\end{align}
Eq. \eqref{eq:Eq105} shows that the impulse fluidity, $\phi(t)$, of the generalized Kelvin-Voigt model is merely the fractional integral of order $p$ (see Eq. \eqref{eq:Eq03}) of the function given by Eq. \eqref{eq:Eq70} $(r=q-p)$
\begin{align}\label{eq:Eq106}
\phi(t)=&\frac{\text{1}}{K_q} I^p \left[ \frac{\text{1}}{t^{\text{1}-p+q}} E_{q-p\text{, }q-p} \left( -\frac{K_p}{K_q} t^{q-p} \right)\right] =  \\ \nonumber
& \frac{\text{1}}{K_q} \frac{\text{1}}{t^{\text{1}-q}} E_{q-p\text{, }q} \left( -\frac{K_p}{K_q} t^{q-p} \right)
\end{align}
where the right-hand side of Eq. \eqref{eq:Eq106} was evaluated by using the general result offered by Eq. \eqref{eq:Eq58} with $\alpha=\beta=q-p$. 

\noindent \textbf{Special Cases:} 1. \textit{Spring -- Scott-Blair element in par\-allel} $(p=\text{0}$, $q \in \mathbb{R}^+ \text{, } K_p=G)$. In this case, Eq. \eqref{eq:Eq106} for $p=$ 0 gives 
\begin{equation}\label{eq:Eq107}
\phi(t)=\frac{\text{1}}{K_q} \frac{\text{1}}{t^{\text{1}-q}} E_{q\text{, }q} \left( -\frac{G}{K_q} t^{q} \right) \text{,} \enskip q \in \mathbb{R}^+
\end{equation}
At the same time, the reader recognizes that for $p=\text{0}$, the coefficient of the fractional integral given by Eq. \eqref{eq:Eq105} vanishes since {\large $\frac{\text{1}}{\Gamma(\text{0})}$} $=\text{0}$. Nevertheless,  the first term under the integral $\left[ \Gamma(p)(t-\xi) \right]^{-\text{1}}$ is the definition of the Dirac delta function given by Eq. \eqref{eq:Eq31} \citep{GelfandShilov1964}. Accordingly, Eq. \eqref{eq:Eq105} reduces to
\begin{align}\label{eq:Eq108}
\phi(t)= & \frac{\text{1}}{K_q}\int_{\text{0}}^{t}\delta(\xi-t)\frac{\text{1}}{\xi^{\text{1}-q}}E_{q\text{, }q}\left( -\frac{G}{K_q}\xi^q \right) \mathrm{d}\xi= \\ \nonumber
&\frac{\text{1}}{K_q} \frac{\text{1}}{t^{\text{1}-q}}E_{q\text{, }q}\left( -\frac{G}{K_q}t^q \right) \text{,} \enskip q \in \mathbb{R}^+
\end{align}
which is the same result offered by Eq. \eqref{eq:Eq107}. When $q=$ 1, $K_q=K_{\text{1}}=\eta$, the impulse fluidity given by Eq. \eqref{eq:Eq108} gives $\phi(t)=$ {\large $\frac{\text{1}}{\eta}$}$E_{\text{1, 1}}\Big( -${\large $\frac{G}{\eta}$}$t \Big)$. Using the identity of the Mittag-Leffler function that $E_{\text{1, 1}}(z)=E_{\text{1}}(z)=e^z$, together with {\large $\nicefrac{\eta}{G}$} $=\lambda=$ relaxation time, $\phi(t)=$ {\large $\frac{\text{1}}{\eta}$}$e${\large $^{-\nicefrac{t}{\lambda}}$} which is the impulse fluidity of the classical Kelvin-Voigt solid shown in Table \ref{tab:Table2}. When $q=$ 2, $K_q=K_{\text{2}}=m_R$ is the distributed inertance of an inerter connected in parallel with a spring with elastic constant $G$. In this case Eq. \eqref{eq:Eq108} gives $\phi(t)=$ {\large $\frac{\text{1}}{m_R}$}$t E_{\text{2, 2}}\Big( -${\large $\frac{G}{m_R}$}$t^{\text{2}} \Big)$. By using that {\large $\frac{G}{m_R}$} $=\omega_R^{\text{2}}$, where $\omega_R$ is the rotational frequency of a spring -- inerter parellel connection together with the identity given by Eq. \eqref{eq:Eq79}, $\phi(t)=$ {\large $\frac{\text{1}}{m_R \omega_R}$}$\sin(\omega_R t)$, which is the impulse fluidity of a spring -- inerter parallel connection \citep{Makris2017}. Eq. \eqref{eq:Eq108} is re-written in its dimensionless form
\begin{equation}\label{eq:Eq109}
\left[  \frac{K_q}{G}\right]^{\nicefrac{\text{1}}{q}} G \phi(t)  = \left[  \frac{G}{K_q} t^q\right]^{\text{1}-\nicefrac{\text{1}}{q}}E_{q\text{, }q}\left( -\frac{G}{K_q} t^q \right) \text{,} \enskip q \in \mathbb{R}^+
\end{equation}
The plots of the right-hand side of Eq. \eqref{eq:Eq109} with a negative sign are depicted in Figure \ref{fig:Fig04} (left) for values of $q=$ 0.5, 1, 1.5, 1.8 and 2 as a function of the dimensionless time {\large $\Big[\frac{G}{K_q}$}$t^q${\large $\Big]^{\nicefrac{\text{1}}{q}}$}.

\begin{figure*}[t!]
\centering
\includegraphics[width=0.9\textwidth, angle=0]{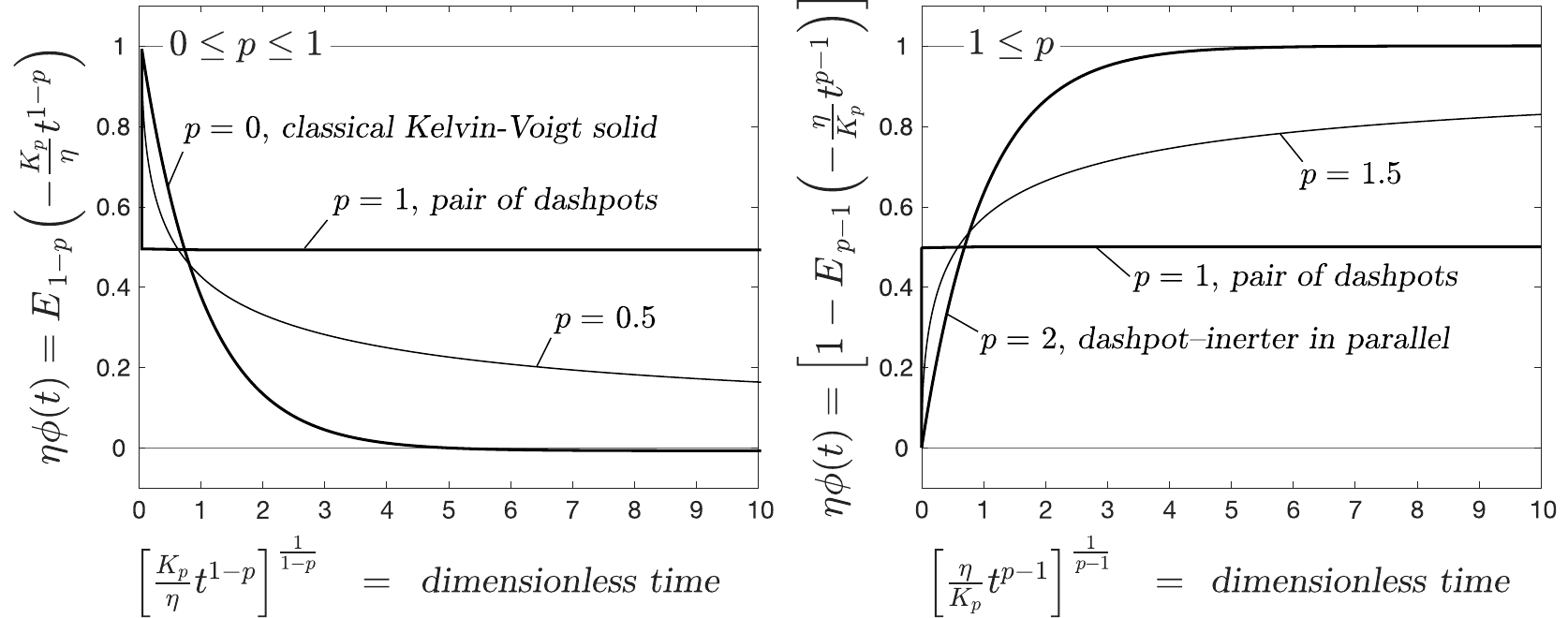}
\caption{Normalized impulse fluidity (impulse response function) of the Scott-Blair -- dashpot parallel connection. The left plots are for $\text{0} \leq p \leq \text{1} $ (springpot -- dashpot in parallel), whereas the right plots are for the Scott-Blair -- dashpot fluid with $p>$ 1.}
\label{fig:Fig06}
\end{figure*}

\noindent 2. \textit{Scott-Blair -- Dashpot in parallel} $(p \in \mathbb{R}^+$, $q=\text{1}$, $K_q=\eta)$. In this case where $q=\text{1}$, the fractional integral of Eq. \eqref{eq:Eq106} gives
\begin{align}\label{eq:Eq110}
\phi(t)=&\frac{\text{1}}{\eta} I^p \left[ \frac{\text{1}}{t^p} E_{\text{1}-p\text{, }\text{1}-p}\left( -\frac{K_p}{\eta}t^{\text{1}-p} \right)  \right] = \\ \nonumber
&\frac{\text{1}}{\eta} E_{\text{1}-p}\left( -\frac{K_p}{\eta}t^{\text{1}-p} \right) 
\end{align}
The result of Eq. \eqref{eq:Eq110} is valid only for 0 $\leq p \leq$ 1 (springpot -- dashpot connection in parallel). When $p=$ 1, $K_p=K_{\text{1}}=\eta_{\text{1}}$ and according to Eq. \eqref{eq:Eq110}, $\phi(t)=$  {\large $\frac{\text{1}}{\eta}$}$E_{\text{0}}$ {\large $\left(- \frac{\eta_{\text{1}}}{\eta} \right)$}, which in association with the identity $E_{\text{0}}(-z)=$ {\large $\frac{\text{1}}{\text{1}+z}$}, Eq. \eqref{eq:Eq110} yields $\phi(t)=$  {\large $\frac{\text{1}}{\eta+\eta_{\text{1}}}$}$U(t-\text{0})$, which is the impulse fluidity of two dashpots connected in parallel. When $p>$ 1 the impulse fluidity of the Scott-Blair element -- dashpot parallel connection can be obtained by returning to Eq. \eqref{eq:Eq103} and setting $q=$ 1 and $K_q=\eta$
\begin{equation}\label{eq:Eq111}
\mathcal{J}(s)=\frac{\text{1}}{K_p} \frac{\text{1}}{s\left( s^{p-\text{1}} +\nicefrac{\eta}{K_p} \right)} = \frac{\text{1}}{\eta} \frac{\nicefrac{\eta}{K_p}}{s\left( s^{p-\text{1}} +\nicefrac{\eta}{K_p} \right)}
\end{equation}
For $p-\text{1}\geq$ 0, the inverse Laplace transform of Eq. \eqref{eq:Eq111} is known
\begin{align}\label{eq:Eq112}
\phi(t)=&\mathcal{L}^{-\text{1}} \left\lbrace \mathcal{J}(s) \right\rbrace = \mathcal{L}^{-\text{1}} \left\lbrace \frac{\text{1}}{\eta} \frac{\nicefrac{\eta}{K_p}}{s\left( s^{p-\text{1}} +\nicefrac{\eta}{K_p} \right)} \right\rbrace = \\ \nonumber
& \frac{\text{1}}{\eta}\left[\text{1}-E_{p-\text{1}}\left( -\frac{\eta}{K_p}t^{p-\text{1}} \right) \right] \text{,} \enskip \text{1}\leq p \in \mathbb{R}
\end{align}
Eq. \eqref{eq:Eq112} offers the impulse fluidity of the Scott-Blair element -- dashpot parallel connection when $p\geq$ 1. When $p=$ 1, $K_p=K_{\text{1}} =\eta_{\text{1}}$ and according to Eq. \eqref{eq:Eq112}, $\phi(t)=$ {\large $\frac{\text{1}}{\eta}$}$\Big[\text{1}-E_{\text{0}}\Big( -${\large $\frac{\eta}{\eta_{\text{1}}}$}$\Big)\Big]$ which in association with the identity $E_{\text{0}}(-z)=$ {\large $\frac{\text{1}}{\text{1}+z}$}, Eq. \eqref{eq:Eq112} yields $\phi(t)=$ {\large $\frac{\text{1}}{\eta+\eta_{\text{1}}}$}$U(t-\text{0})$, which is the result of Eq. \eqref{eq:Eq110} for $p=$ 1 and the continuity of the two solutions is established. When $p=$ 2, $K_p=K_{\text{2}}=m_R$, $\phi(t)=$ {\large $\frac{\text{1}}{\eta}$}$\Big[\text{1}-E_{\text{1}}\Big( -${\large $\frac{\eta}{m_R}$}$t\Big)\Big]$ $=$ {\large $\frac{\text{1}}{\eta}\Big[$}$U(t-\text{0}) -e${\large $^{-\frac{\eta}{m_R}t} \Big]$}, which is the impulse fluidity of a dashpot -- inerter parallel connection \citep{Makris2017}.

Figure \ref{fig:Fig06} plots the dimensionless impulse fluidity, $\eta \phi(t)$, of the Scott-Blair -- dashpot parallel connection for values of $p=$ 0, 0.5, 1, 1.5 and 2 by employing Eq. \eqref{eq:Eq110} when $\text{0} \leq p \leq \text{1}$ and Eq. \eqref{eq:Eq112} for $\text{1} \leq p$.

Given the parallel connection of the two Scott-Blair elements, the relaxation modulus, $G(t)$, of the generalized fractional Kelvin-Voigt element is the summation of the relaxation moduli of the two individual Scott-Blair elements given by Eq. \eqref{eq:Eq49}

\begin{equation}\label{eq:Eq113}
G(t)=\left[ \frac{K_p}{\Gamma(\text{1}-p)}\frac{\text{1}}{t^p} + \frac{K_q}{\Gamma(\text{1}-q)}\frac{\text{1}}{t^q} \right] U(t-\text{0})
\end{equation}

\noindent \textbf{Special Cases:} 1. \textit{Spring -- Scott-Blair element in parallel} $(p=\text{0}$, $q \in \mathbb{R}^+ \text{, } K_p=G)$
\begin{equation}\label{eq:Eq114}
G(t)=\left[ G + K_q\frac{\text{1}}{\Gamma(\text{1}-q)}\frac{\text{1}}{t^q} \right]U(t-\text{0})
\end{equation}
The result of Eq. \eqref{eq:Eq114} has been first presented by \citet{Koeller1984}.

\noindent 2. \textit{Springpot -- Dashpot in parallel} $(q=\text{1}$, $K_q=\eta)$
\begin{equation}\label{eq:Eq115}
G(t)= K_p\frac{\text{1}}{\Gamma(\text{1}-p)}\frac{\text{1}}{t^p} U(t-\text{0}) +\eta \delta(t-\text{0})
\end{equation}
The Dirac delta function in the right-hand side of Eq. \eqref{eq:Eq115} emerges by virtue of Eq. \eqref{eq:Eq31} for $n=\text{0}$.

The complex dynamic fluidity (admittance), $\phi(s)$, of the generalized fractional Kelvin-Voigt element derives directly from Eq. \eqref{eq:Eq103} by using that $\dt{\gamma}(s)=s\gamma(s)$
\begin{align}\label{eq:Eq116}
\phi(s)=&\frac{\text{1}}{K_q}\frac{s}{s^p}\frac{\text{1}}{s^{q-p}+\frac{K_p}{K_q}}= \\ \nonumber
& \frac{\text{1}}{K_q}s^{\text{1}-p}\frac{\text{1}}{s^r+\frac{K_p}{K_q}} \text{,} \enskip r=q-p>\text{0}
\end{align}
The inverse Laplace transform of Eq. \eqref{eq:Eq116} is evaluated with the convolution integral given by Eq. \eqref{eq:Eq68}, where according to Eq. \eqref{eq:Eq40}
\begin{align}\label{eq:Eq117}
f(t)=&\mathcal{L}^{-\text{1}} \left\lbrace \mathcal{F}(s) \right\rbrace = \mathcal{L}^{-\text{1}} \left\lbrace \frac{\text{1}}{K_q}s^{\text{1}-p} \right\rbrace = \\ \nonumber
& \frac{\text{1}}{K_q} \frac{\text{1}}{\Gamma(-\text{1}+p)} \frac{\text{1}}{t^{\text{2}-p}}U(t-\text{0}) \text{,} \enskip \text{0}<p \in \mathbb{R}^+
\end{align}
and $h(t)=\mathcal{L}^{-\text{1}} \left\lbrace \mathcal{H}(s) \right\rbrace $ is given by Eq. \eqref{eq:Eq70}. Substitution of the results of Eqs. \eqref{eq:Eq117} and \eqref{eq:Eq70} into the convolution integral given by Eq. \eqref{eq:Eq68}, the impulse strain-rate response function of the generalized fractional Kelvin-Voigt element is
\begin{align}\label{eq:Eq118}
\psi(t)&=\mathcal{L}^{-\text{1}} \left\lbrace \phi(s) \right\rbrace =  \\ \nonumber
&\frac{\text{1}}{K_q} \frac{\text{1}}{\Gamma(\text{1}-p)}\int_{\text{0}}^{t} \frac{\text{1}}{(t-\xi)^{\text{2}-p}} \xi^{r-\text{1}} E_{r\text{, }r}\left( -\frac{K_p}{K_q} \xi^r \right) \mathrm{d}\xi
\end{align}
Eq. \eqref{eq:Eq118} shows that the impulse strain-rate function, $\psi(t)$, of the generalized Kelvin-Voigt element is merely the fractional derivative of order $\text{1}-p$ (see Eq. \eqref{eq:Eq04}) of the function given by Eq. \eqref{eq:Eq70}. After replacing $r$ with $q-p$, Eq. \eqref{eq:Eq118} gives
\begin{align}\label{eq:Eq119}
\psi(t)=&\frac{\text{1}}{K_q} \frac{\mathrm{d}^{\text{1}-p}}{\mathrm{d}t^{\text{1}-p}} \left[ \frac{\text{1}}{t^{\text{1}-q+p}} E_{q-p\text{, }q-p}\left( -\frac{K_p}{K_q}t^{q-p} \right) \right] = \\ \nonumber
&  \frac{\text{1}}{K_q} t^{q-\text{2}} E_{q-p\text{, }q-\text{1}}\left( -\frac{K_p}{K_q}t^{q-p} \right)  \text{,} \enskip \text{0}<p<q \in \mathbb{R}^+
\end{align}
where the right-hand side of Eq. \eqref{eq:Eq119} was obtained by using the result of Eq. \eqref{eq:Eq59} where $\beta=q-p$. The right-hand side of Eq. \eqref{eq:Eq119} is valid for $q>$ 1. For the case where 0 $\leq q \leq$ 1 (springpot -- Scott-Blair element in parallel), the singularity embedded in the impulse strain-rate response function, $\psi(t)$, of the generalized fractional Kelvin-Voigt element is extracted by virtue of Eq. \eqref{eq:Eq61} with $\beta=q-p$
\begin{align}\label{eq:Eq120}
\psi(t)=&\frac{\text{1}}{K_q} \Bigg[   \frac{\mathrm{d}^{\text{1}-q}}{\mathrm{d}t^{\text{1}-q}}\delta(t-\text{0}) - \\ \nonumber
& \frac{K_p}{K_q} \frac{\text{1}}{t^{\text{2}-\text{2}q+p}} E_{q-p\text{, }\text{2}q-p-\text{1}}\left(- \frac{K_p}{K_q} t^{q-p} \right) \Bigg]
\end{align}
In the event that $\text{2}q-p-\text{1}$ remains negative $(\text{2}q-p-\text{1}<\text{0})$ application once again of the recurrence relation \eqref{eq:Eq57} on the MIttag-Leffler function appearing on the right-hand side of Eq. \eqref{eq:Eq120} gives
\begin{align}\label{eq:Eq121}
\psi(t)=&\frac{\text{1}}{K_q} \Bigg[   \frac{\mathrm{d}^{\text{1}-q}}{\mathrm{d}t^{\text{1}-q}}\delta(t-\text{0}) - \frac{K_p}{K_q}  \frac{\mathrm{d}^{\text{1}-\text{2}q+p}}{\mathrm{d}t^{\text{1}-\text{2}q+p}}\delta(t-\text{0}) + \\ \nonumber
&\left( \frac{K_p}{K_q} \right)^{\text{2}}
\frac{\text{1}}{t^{\text{2}-\text{3}q+\text{2}p}} E_{q-p\text{, }\text{3}q-\text{2}p-\text{1}}\left(- \frac{K_p}{K_q} t^{q-p} \right) \Bigg]
\end{align}
where now the second singularity {\large $\frac{\mathrm{d}^{\text{1}-\text{2}q+p}}{\mathrm{d}t^{\text{1}-\text{2}q+p}}$}$\delta(t-\text{0})$ has been extracted. In the event that $\text{3}q-\text{2}p-\text{1}$ remains negative, this procedure can be repeated until the second index of the MIttag-Leffler function appearing on the right-hand side of the impulse strain-rate response function, $\psi(t)$, is positive and all the singularities will have been extracted.
\begin{figure*}[!]
\centering
\includegraphics[width=0.9\textwidth, angle=0]{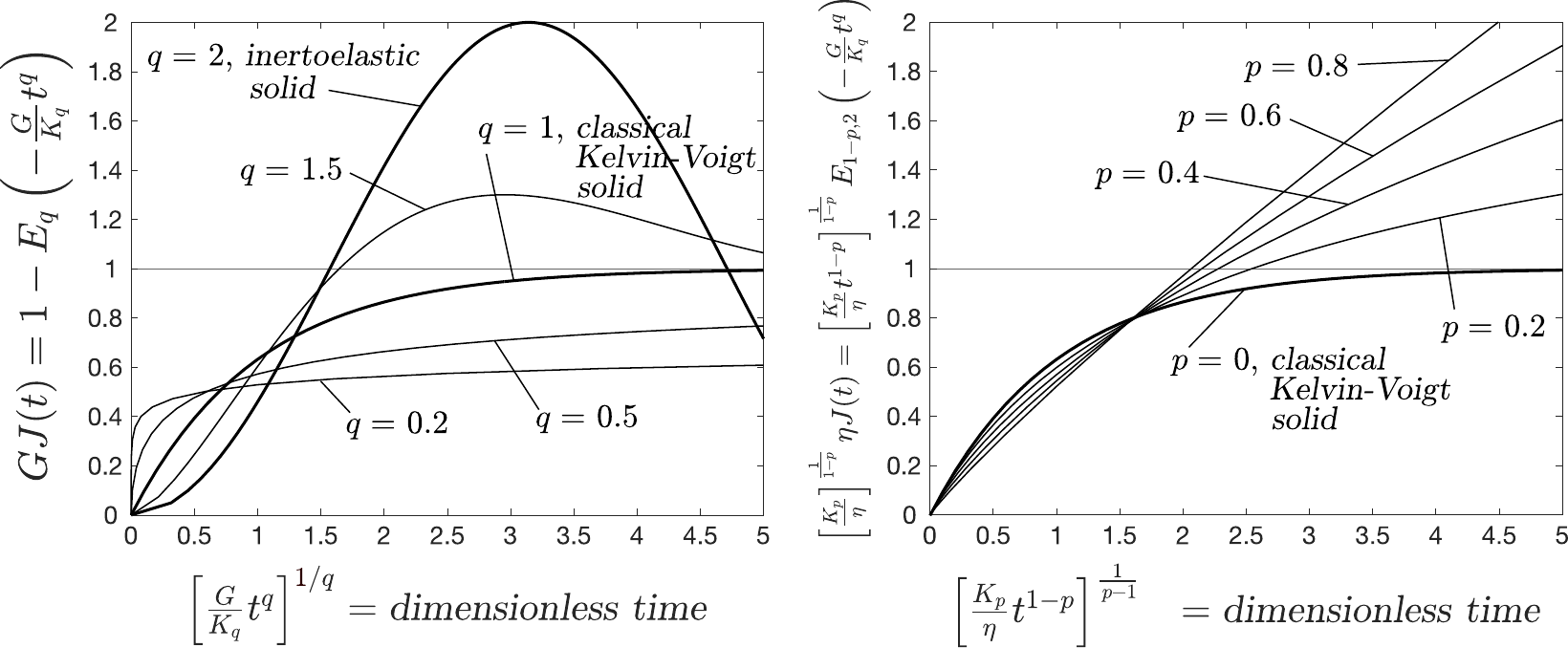}
\caption{Left: Normalized creep compliance, $GJ(t)$, of the spring -- Scott-Blair element in parallel for values $q=$ 0.2, 0.5, 1, 1.5 and 2 as a function of the dimensionless time {\large $\Big[\frac{G}{K_q}$}$t^q${\large $\Big]^{\nicefrac{\text{1}}{q}}$}. Right: Normalized creep compliance, {\large $\left[\frac{K_p}{\eta}\right]^{\frac{\text{1}}{\text{1}-p}}$}$\eta J(t)$, of the springpot -- dashpot element in parallel for values of $p=$ 0, 0.2, 0.4, 0.6 and 0.8 as a function of the dimensionless time {\large $\Big[\frac{K_p}{\eta}$}$t^{\text{1}-p}${\large $\Big]^{\frac{\text{1}}{\text{1}-p}}$}.}
\label{fig:Fig07}
\end{figure*}

\noindent \textbf{Special Cases:} 1. \textit{Spring -- Scott-Blair element in parallel} $(p=\text{0}$, $q \in \mathbb{R}^+ \text{, } K_p=G)$
\begin{align}\label{eq:Eq122}
\psi(t)=&\frac{\text{1}}{K_q} \frac{\mathrm{d}}{\mathrm{d}t} \left[ \frac{\text{1}}{t^{\text{1}-q}} E_{q\text{, }q}\left( -\frac{G}{K_q}t^q \right) \right] =  \\ \nonumber
&\frac{\text{1}}{K_q}  t^{q-\text{2}} E_{q\text{, }q-\text{1}}\left( -\frac{G}{K_q}t^q \right) 
\end{align}
Again the right-hand side of Eq. \eqref{eq:Eq122} is valid for $q>$ 1. For the case when 0 $\leq q \leq$ 1 (spring -- springpot in parallel) we use Eq. \eqref{eq:Eq120} with $p=$ 0
\begin{align}\label{eq:Eq123}
\psi(t)=&\frac{\text{1}}{K_q} \Bigg[   \frac{\mathrm{d}^{\text{1}-q}}{\mathrm{d}t^{\text{1}-q}}\delta(t-\text{0}) -  \\ \nonumber
& \frac{G}{K_q} \frac{\text{1}}{t^{\text{2}-\text{2}q}} E_{q\text{, }\text{2}q-\text{1}}\left(- \frac{G}{K_q} t^{q} \right) \Bigg]
\end{align}
In the event that $q\leq$ 0.5 $(\text{2}q-\text{1}\leq \text{0})$, the expression for $\psi(t)$ offered by Eq. \eqref{eq:Eq121} needs to be used with $p=$ 0.

\noindent 2. \textit{Springpot -- Dashpot in parallel} $(q=\text{1}$, $K_q=\eta$, $\text{0} \leq p \leq \text{1})$. In this case we need to use directly Eq. \eqref{eq:Eq120} with $p=$ 1
\begin{equation}\label{eq:Eq124}
\psi(t)= \frac{\text{1}}{\eta} \left[ \delta(t-\text{0}) -  \frac{K_p}{\eta} \frac{\text{1}}{t^p} E_{\text{1}-p\text{, }\text{1}-p}\left( -\frac{K_p}{\eta}t^{\text{1}-p} \right) \right]
\end{equation}

The complex creep function, $\mathcal{C}(\omega)$,  of the generalized fractional Kelvin-Voigt element derives directly from Eq. \eqref{eq:Eq103} gives that $\mathcal{C}(s)=$ {\large $\nicefrac{\mathcal{J}(s)}{s}$}
\begin{equation}\label{eq:Eq125}
\mathcal{C}(s)= \frac{\gamma(s)}{\dt{\tau}(s)}=\frac{\text{1}}{K_q} \frac{\text{1}}{s^{\text{1}+p}(s^r+\frac{K_p}{K_q})} \text{,} \enskip \text{with} \enskip p\text{, }r \in \mathbb{R}^+
\end{equation}
and $r=q-p> \text{0}$.

The inverse Laplace transform of Eq. \eqref{eq:Eq125} is evaluated with the convolution integral given by Eq. \eqref{eq:Eq68} where
\begin{align}\label{eq:Eq126}
f(t)=&\mathcal{L}^{-\text{1}} \left\lbrace \mathcal{F}(s) \right\rbrace = \mathcal{L}^{-\text{1}} \left\lbrace \frac{\text{1}}{K_q} \frac{\text{1}}{s^{\text{1}+p}} \right\rbrace = \\ \nonumber
&  \frac{\text{1}}{K_q} \frac{\text{1}}{\Gamma(\text{1}+p)}t^p \text{,} \enskip p \in \mathbb{R}^+
\end{align}
and $h(t)=\mathcal{L}^{-\text{1}} \left\lbrace \mathcal{H}(s) \right\rbrace $ is given by Eq. \eqref{eq:Eq70}. Substitution of the results of Eqs. \eqref{eq:Eq126} and \eqref{eq:Eq70} into the convolution integral given by Eq. \eqref{eq:Eq68}, the creep compliance, $J(t)$, of the generalized fractional Kelvin-Voigt element is
\begin{align}\label{eq:Eq127}
J(t)=&\mathcal{L}^{-\text{1}} \left\lbrace \mathcal{C}(s) \right\rbrace = \\ \nonumber
&\frac{\text{1}}{K_q} \frac{\text{1}}{\Gamma(\text{1}+p)} \int_{\text{0}}^{t} (t-\xi)^p \xi^{r-\text{1}} E_{r\text{, }r}\left( -\frac{K_p}{K_q}\xi^r\right) \mathrm{d}\xi  
\end{align}
which is the fractional integral of orde $\text{1}+p$ of the function given by Eq. \eqref{eq:Eq70} in which $r=q-p$
\begin{align}\label{eq:Eq128}
J(t)=&\frac{\text{1}}{K_q}I^{\text{1}+p} \left[ \frac{\text{1}}{t^{\text{1}-q+p}} E_{q-p\text{, }q-p} \left( -\frac{K_p}{K_q}t^{q-p} \right) \right]= \\ \nonumber
& \frac{\text{1}}{K_q}t^q E_{q-p\text{, }q+\text{1}} \left( -\frac{K_p}{K_q}t^{q-p} \right) \text{,} \enskip \text{0}<q<p \in \mathbb{R}^+
\end{align}
where the right-hand side of Eq. \eqref{eq:Eq128} was evaluated by using the general result offered by Eq. \eqref{eq:Eq58} with $\alpha=\beta=q-p$. The result of Eq. \eqref{eq:Eq128} has been presented by \citet{SchiesselMetzlerBlumenNonnenmacher1995, Hristov2019}.

\begin{sidewaystable*}
\centering
\caption{Frequency-response functions and the corresponding causal time-response functions of the generalized fractional derivative Kelvin-Voigt element and of its special cases.}
\setlength{\tabcolsep}{2pt}
{\renewcommand{\arraystretch}{1.5}
\begin{tabularx}{\textheight}{>{\centering}p{0.13\textheight}>{\centering}p{0.29\textheight}>{\centering}p{0.22\textheight}>{\centering}p{0.36\textheight}}
\hline \hline
	& \thead{\textbf{Generalized Fractional Derivative} \\ \textbf{ Kelvin-Voigt Element}} & \thead{\textbf{Spring -- Scott-Blair} \\ \textbf{Parallel Connection}} & \thead{\textbf{Springpot -- Dashpot} \\ \textbf{Parallel Connection}} \tabularnewline
	& \includegraphics[scale=0.25]{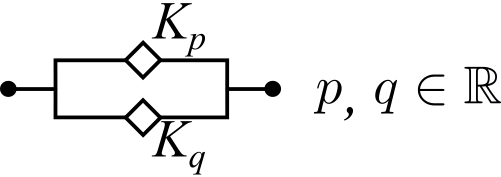}  & \includegraphics[scale=0.25]{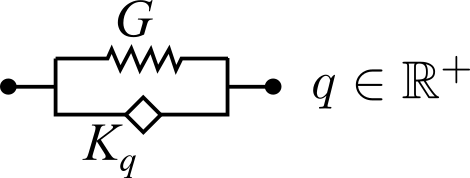} & \includegraphics[scale=0.25]{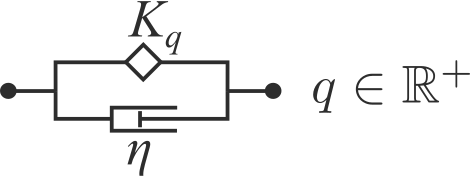}
	\tabularnewline
	\thead{Constitutive \\ Equation} & $\tau(t)=K_p\cfrac{\mathrm{d}^p\gamma(t)}{\mathrm{d}t^p}+K_q\cfrac{\mathrm{d}^q\gamma(t)}{\mathrm{d}t^q}$  &  $\tau(t)=G\gamma(t)+ K_q\cfrac{\mathrm{d}^q\gamma(t)}{\mathrm{d}t^q}$  & $\tau(t)=K_q \cfrac{\mathrm{d}^q\gamma(t)}{\mathrm{d}t^q}+\eta \cfrac{\mathrm{d}\gamma(t)}{\mathrm{d}t}$
	\tabularnewline \hline
	\thead{Complex \\ Dynamic Modulus \\ $\mathcal{G}(\omega)=\cfrac{\tau(\omega)}{\gamma(\omega)}$}  & {\normalsize $K_p(\operatorname{i}\omega)^p + K_q(\operatorname{i}\omega)^q$} & {\normalsize $G+K_q(\operatorname{i}\omega)^q$ }& {\normalsize $K_q(\operatorname{i}\omega)^q+\eta\operatorname{i}\omega$}
	\tabularnewline
	\thead{Complex\\ Dynamic Viscosity \\ $\mathcal{\eta}(\omega)=\frac{\tau(\omega)}{\dt{\gamma}(\omega)}$}  & $\cfrac{K_p(\operatorname{i}\omega)^p + K_q(\operatorname{i}\omega)^q}{\operatorname{i}\omega}$ &  $G \left[ \pi \delta(\omega-\text{0})-\operatorname{i}\cfrac{\text{1}}{\omega} \right]+K_q \cfrac{\text{1}}{(\operatorname{i}\omega)^{\text{1}-q}}$ & $ \eta +K_q\cfrac{\text{1}}{(\operatorname{i}\omega)^{\text{1}-q}}$
	\tabularnewline 
	\thead{Complex \\ Dynamic Compliance \\ $\mathcal{J}(\omega)=\frac{\text{1}}{\mathcal{G}(\omega)}=\frac{\gamma(\omega)}{\tau(\omega)}$} & $\cfrac{\text{1}}{K_p(\operatorname{i}\omega)^p + K_q(\operatorname{i}\omega)^q}$ & $\cfrac{\text{1}}{G+K_q(\operatorname{i}\omega)^q}$ & $\cfrac{\text{1}}{\eta}\cfrac{\eta\big/K_q}{\operatorname{i}\omega\left[ (\operatorname{i}\omega)^{-\text{1}+q} +\frac{\eta}{K_q} \right]}$
	\tabularnewline
	\thead{Complex  \\ Creep Function \\ $\mathcal{C}(\omega)=\frac{\gamma(\omega)}{\dt{\tau}(\omega)}$}  & $\cfrac{\text{1}}{K_p(\operatorname{i}\omega)^{p+\text{1}} + K_q(\operatorname{i}\omega)^{q+\text{1}}}$  & $\cfrac{\text{1}}{\operatorname{i}\omega\left[ G+ K_q (\operatorname{i}\omega)^q \right]}$ &  $\cfrac{\text{1}}{\eta}\cfrac{\eta\big/K_q}{\operatorname{i}\omega\left[ (\operatorname{i}\omega)^q +\frac{\eta}{K_q}\operatorname{i}\omega \right]}$
	\tabularnewline 
	\thead{ Complex \\ Dynamic Fluidity\\ $\mathcal{\phi}(\omega)=\frac{\text{1}}{\eta(\omega)}=\frac{\dt{\gamma}(\omega)}{\tau(\omega)}$} & $\cfrac{\operatorname{i}\omega}{K_p(\operatorname{i}\omega)^p + K_q(\operatorname{i}\omega)^q}$ &  $\cfrac{\operatorname{i}\omega}{ G+ K_q (\operatorname{i}\omega)^q}$ & $\cfrac{\text{1}}{\eta}\cfrac{(\operatorname{i}\omega)^{\text{1}-q}}{(\operatorname{i}\omega)^{\text{1}-q}+\frac{K_q}{\eta}}$
	\tabularnewline \hline
	\thead{Memory Function \\ $M(t)$}	 & $\left[ \cfrac{K_p}{\Gamma(-p)} \cfrac{\text{1}}{t^{p+\text{1}}} + \cfrac{K_q}{\Gamma(-q)} \cfrac{\text{1}}{t^{q+\text{1}}} \right]U(t-\text{0})$ & $G \delta(t-\text{0}) +\cfrac{K_q}{\Gamma(-q)}\cfrac{\text{1}}{t^{q+1}}U(t-\text{0})$ & $\cfrac{K_q}{\Gamma(-q)}\cfrac{\text{1}}{t^{q+\text{1}}}U(t-\text{0}) + \eta\cfrac{\mathrm{d}\delta(t-\text{0})}{\mathrm{d}t}$
	\tabularnewline 
	\thead{Relaxation Modulus \\ $G(t)$} & $\left[ \cfrac{K_p}{\Gamma(\text{1}-p)} \cfrac{\text{1}}{t^p} + \cfrac{K_q}{\Gamma(\text{1}-q)} \cfrac{\text{1}}{t^q} \right]U(t-\text{0})$ & $\left[G +\cfrac{K_q}{\Gamma(\text{1}-q)}\cfrac{\text{1}}{t^q}\right]U(t-\text{0})$ & $\cfrac{K_q}{\Gamma(\text{1}-q)}\cfrac{\text{1}}{t^q}U(t-\text{0}) + \eta \delta(t-\text{0})$
	\tabularnewline 
	\thead{Impulse Fluidity \\ $\phi(t)$}   & $\cfrac{\text{1}}{K_q}  \cfrac{\text{1}}{t^{\text{1}-q}} E_{q-p\text{, }q} \left( -\cfrac{K_p}{K_q} t^{q-p} \right) $ & $\cfrac{\text{1}}{K_q}\cfrac{\text{1}}{t^{\text{1}-q}}E_{q\text{, }q}\left( -\cfrac{G}{K_q}t^q \right)$ & \thead{{\scriptsize $\cfrac{\text{1}}{\eta}\left[ E_{\text{1}-q} \left( -\cfrac{K_q}{\eta}\,t^{\text{1}-q} \right) \right]$}\\ {\scriptsize 0 $\leq q \leq$ 1}}\textbf{or}\thead{{\scriptsize$\cfrac{\text{1}}{\eta}\left[ \text{1} - E_{q-\text{1}} \left( -\cfrac{\eta}{K_q}\,t^{q-\text{1}} \right) \right]$}\\ {\scriptsize 1 $\leq q \in \mathbb{R}^+$}}
	\tabularnewline
	\thead{Creep Compliance \\ $J(t)$}	& $\cfrac{\text{1}}{K_q}t^q E_{q-p\text{, }q+\text{1}} \left( -\cfrac{K_p}{K_q}t^{q-p} \right) $ & $\cfrac{\text{1}}{G}\left[ U(t-\text{0}) - E_q\left( -\cfrac{G}{K_q}t^q \right) \right]$ & $\cfrac{\text{1}}{\eta} t  E_{\text{1}-q\text{, 2}} \left( -\cfrac{K_q}{\eta} t^{\text{1}-q} \right)  $
	\tabularnewline 
	\thead{Impulse Strain-rate \\ Response Function \\ $\psi(t)$}	& \thead{$\cfrac{\text{1}}{K_q}\cfrac{\text{1}}{t^{\text{2}-q}} \Bigg[ \cfrac{\text{1}}{\Gamma(-\text{1}+q)}\,U(t-\text{0}) - \quad \quad \quad  $ \\ $  \quad \quad \quad  \cfrac{K_p}{K_q}  \cfrac{\text{1}}{t^{p-q}} E_{q-p\text{, }\text{2}q-p-\text{1}}\left( -\cfrac{K_p}{K_q}t^{q-p} \right) \Bigg] $} & \thead{$\cfrac{\text{1}}{K_q}\cfrac{\text{1}}{t^{\text{2}-q}} \Bigg[ \cfrac{\text{1}}{\Gamma(-\text{1}+q)}\,U(t-\text{0}) - \quad  $ \\ $ \quad   \cfrac{G}{K_q}  \cfrac{\text{1}}{t^{-q}} E_{q\text{, }\text{2}q-\text{1}}\left( -\cfrac{G}{K_q}t^{q} \right) \Bigg] $} & $\cfrac{\text{1}}{\eta} \Bigg[ \delta(t-\text{0}) - \cfrac{K_q}{\eta} \cfrac{\text{1}}{t^q} E_{\text{1}-q\text{, }\text{1}-q} \left( -\cfrac{K_q}{\eta} t^{\text{1}-q} \right) \Bigg] $
	\tabularnewline \hline \hline
\end{tabularx}}
\label{tab:Table4}
\end{sidewaystable*}

\noindent \textbf{Special Cases:} 1. \textit{Spring -- Scott-Blair element in parallel} $(p=\text{0}$, $K_p=G\text{, } q \in \mathbb{R}^+  )$. In this case where $p=\text{0}$, Eq. \eqref{eq:Eq128} gives
\begin{equation}\label{eq:Eq129}
J(t)=\frac{\text{1}}{K_q}t^q E_{q\text{, }q+\text{1}} \left( -\frac{G}{K_q}t^{q} \right)
\end{equation}
Alternatively, the creep compliance, $J(t)$, for $p=$ 0 can be evaluated by returning to the expression of the complex creep function, $\mathcal{C}(s)$, of the generalized fractional Kelvin-Voigt element given by Eq. \eqref{eq:Eq125} by examining the special case where $r=q-p=q$
\begin{equation}\label{eq:Eq130}
\mathcal{C}(s)=\frac{\text{1}}{K_q}\frac{\text{1}}{s(s^q+\frac{G}{K_q})}=\frac{\text{1}}{G} \frac{\frac{G}{K_q}}{s(s^q+\frac{G}{K_q})}
\end{equation}
The inverse Laplace transform of the right-hand side of Eq. \eqref{eq:Eq130} is known
\begin{align}\label{eq:Eq131}
J(t)=&\mathcal{L}^{-\text{1}} \left\lbrace \mathcal{C}(s) \right\rbrace=\mathcal{L}^{-\text{1}} \left\lbrace \frac{\text{1}}{G} \frac{\frac{G}{K_q}}{s(s^q+\frac{G}{K_q})} \right\rbrace = \\ \nonumber
&\frac{\text{1}}{G} \left[ \text{1} - E_q \left( -\frac{G}{K_q}t^q \right) \right] \text{, } q \in \mathbb{R}^+
\end{align}
The result of Eq. \eqref{eq:Eq131} has been presented by \citet{Koeller1984, SchiesselMetzlerBlumenNonnenmacher1995} in their studies in viscoelasticity and by \citet{WesterlundEkstam1994} for a capacitor model of mixed dielectrics. By employing the recurrence relation \eqref{eq:Eq57} of the Mittag-Leffler function
\begin{align} \label{eq:Eq132}
 E_q \left( -\frac{G}{K_q}t^q \right) = & E_{q\text{, 1}} \left( -\frac{G}{K_q}t^q \right) = \\ \nonumber
&\text{1} - \frac{G}{K_q}t^q E_{q\text{, }q+\text{1}} \left( -\frac{G}{K_q}t^q \right)
\end{align}
the substitution of the right-hand side of Eq. \eqref{eq:Eq132} into the right-hand side of Eq. \eqref{eq:Eq131} yields the expression offered by Eq. \eqref{eq:Eq129}; therefore the result offered by Eq. \eqref{eq:Eq129} (outcome of the fractional integral) and the result offered by Eq. \eqref{eq:Eq131} (inverse Laplace transform of the complex creep function) are identical. Figure \ref{fig:Fig07} (left) plots the normalized creep compliance (retardation function) of the spring -- Scott-Blair element parallel connection, $GJ(t)$, for values of $q=$ 0.2, 0.5, 1, 1.5 and 2 as a function of the dimensionless time {\large $\Big[\frac{G}{K_q}$}$t^q${\large $\Big]^{\nicefrac{\text{1}}{q}}$}. When $q=$ 2 (inertoelastic solid), the creep compliance, $J(t)$, exhibits oscillatory behavior since the spring and the inerter exchange their potential (spring) and kinetic (inerter) energies.

\noindent 2. \textit{Springpot -- Dashpot in parallel} $(q=\text{1}$, $K_q=\eta$, $\text{0}<p<\text{1})$. In this case where $q=\text{1}$ and $K_q=K_{\text{1}}=\eta$, Eq. \eqref{eq:Eq128} yields
\begin{equation}\label{eq:Eq133}
J(t)=\frac{\text{1}}{\eta}t E_{\text{1}-p\text{, }\text{2}} \left( -\frac{K_p}{\eta}t^{\text{1}-p} \right) 
\end{equation}
For the limit case where $p=$ 0 and $K_p=K_{\text{0}}=G$, Eq. \eqref{eq:Eq133} yields $J(t)=\frac{\text{1}}{\eta}t E_{\text{1, 2}}${\large $\Big(-\frac{G}{\eta}$}$t$ {\large $\Big)$}. Using the identity that $E_{\text{1, 2}}(z)=$ {\large $\frac{e^z-\text{1}}{z}$}, together with {\large $\nicefrac{\eta}{G}=$} $\lambda=$ relaxation time, $J(t)=$ {\large $\frac{\text{1}}{G}\Big[$}$U(t-\text{0})-e${\large $^{-\nicefrac{t}{\lambda}}\Big]$}, which is the creep compliance of the classical Kelvin-Voigt solid. Figure \ref{fig:Fig07} (right) plots the normalized creep compliance of the springpot -- dashpot parallel connection,  {\large $\left[\frac{K_p}{\eta}\right]^{\frac{\text{1}}{\text{1}-p}}$}$\eta J(t)$, for various values of $p$.  

The five causal time-response functions of the generalized fractional Kelvin-Voigt element together with the time-response functions for the special cases of the spring -- Scott-Blair element parallel connection $(p=\text{0})$, and the springpot -- dashpot parallel connection $(q=\text{1})$ are summarized in Table \ref{tab:Table4}.

\section{Conclusions}
In this paper we studied the five time-response functions of the generalized fractional derivative Maxwell fluid and the generalized fractional derivative Kelvin-Voigt element. These two rheological models are in-series or parallel connections of two Scott-Blair elements which are the elementary fractional derivative elements. In this work the order of differentiation in each Scott-Blair element is allowed to exceed unity reaching values up to 2; and at this limit case the Scott-Blair element becomes an inerter. With this generalization, where the Scott-Blair element goes beyond the traditional springpot, inertia effects may be captured in addition to the monotonic viscoelastic effects. In the special case of spring -- inerter connections, the time response functions which are not superpositions exhibit oscillatory behavior given the continuous exchange of potential energy (spring) and kinetic energy (inerter).

In addition to the well studied relaxation moduli and creep compliances of the two generalized fractional derivative rheological models, we compute closed-form expressions of the remaining three time-response functions which are the memory function, the impulse fluidity (impulse response function) and the impulse strain-rate response function. Central role to these calculations plays the fractional derivative of the Dirac delta function which is merely the kernel appearing in the convolution of the Riemann-Liouville definition of the fractional derivative of a function and it is the generalization of the \citet{GelfandShilov1964} definition of the Dirac delta function and its integer-order derivatives for any positive real number. This finding shows that the fractional derivative of the Dirac delta function, {\large $\frac{\mathrm{d}^q\delta(t-\text{0})}{\mathrm{d}t^q}$}, is finite everywhere other than at the singularity point and it is the inverse Fourier transform of $(\operatorname{i}\omega)^q$ where $q$ is any positive real number. The fractional derivative of the Dirac delta function emerges as key function in the derivation of the time-response functions of generalized fractional derivative rheological models, since it makes possible the extraction of the singularities embedded in the fractional derivatives of the two-parameter Mittag-Leffler function that emerges invariably in the time-response functions of fractional derivative rheological models. The mathematical techniques developed in this work can be applied to calculate the time-response function of higher-parameter rheological models that involve fractional-order time derivatives.


%
\noindent \\ \textbf{Conflict of interest} The authors declare that they have no conflict of interest.

\noindent \\ \textbf{Funding} Not applicable.

\bibliographystyle{spbasic}      
\bibliography{References}   

\end{document}